\documentclass[iop,numberedappendix]{emulateapj}
\usepackage{epsfig}
\usepackage{amsmath}
\usepackage{enumerate}
\usepackage{natbib}
\usepackage[backref,breaklinks,colorlinks,citecolor=blue]{hyperref}

\usepackage{ulem}
\usepackage{color}

\newbox\grsign \setbox\grsign=\hbox{$>$} \newdimen\grdimen \grdimen=\ht\grsign
\newbox\simlessbox \newbox\simgreatbox \newbox\simpropbox
\setbox\simgreatbox=\hbox{\raise.5ex\hbox{$>$}\llap
  {\lower.5ex\hbox{$\sim$}}}\ht1=\grdimen\dp1=0pt
\setbox\simlessbox=\hbox{\raise.5ex\hbox{$<$}\llap
  {\lower.5ex\hbox{$\sim$}}}\ht2=\grdimen\dp2=0pt
\setbox\simpropbox=\hbox{\raise.5ex\hbox{$\propto$}\llap
  {\lower.5ex\hbox{$\sim$}}}\ht2=\grdimen\dp2=0pt
\def\simgt{\mathrel{\copy\simgreatbox}}
\def\simlt{\mathrel{\copy\simlessbox}}

\newcommand{\be}{\begin{equation}}
\newcommand{\ee}{\end{equation}}

\newcommand{\p}{\partial}
\newcommand{\D}{\mathrm{d}}

\newcommand{\rhop}{\ensuremath{\rho_\mathrm{p}}}
\newcommand{\hs}{\ensuremath{\tilde{h}}}
\newcommand{\sigmat}{\ensuremath{\sigma_\mathrm{T}}}
\newcommand{\xiKN}{\ensuremath{\xi_\mathrm{KN}}}

\shorttitle{RMSs in GRBs: Pair creation}
\shortauthors{Lundman, Beloborodov \& Vurm}

\begin{document}

\title{Radiation mediated shocks in gamma-ray bursts: Pair creation}
\author{Christoffer Lundman\altaffilmark{1,2,3}, Andrei M. Beloborodov\altaffilmark{1} and Indrek Vurm\altaffilmark{1,4}}
\affil{$^1$Physics Department and Columbia Astrophysics Laboratory, Columbia University, 538 West 120th Street, New York, NY 10027, USA \\
$^2$Department of Physics, KTH Royal Institute of Technology, AlbaNova, SE-106 91 Stockholm, Sweden \\
$^3$The Oskar Klein Centre for Cosmoparticle Physics, AlbaNova, SE-106 91 Stockholm, Sweden \\
$^4$Tartu Observatory, T\~{o}ravere 61602, Tartumaa, Estonia}

\label{firstpage}
\begin{abstract}
Sub-photospheric shock dissipation is one of the main proposed mechanisms for producing the prompt gamma-ray burst (GRB) emission. Such shocks are mediated by scattering of radiation. We introduce a time dependent, special relativistic code which dynamically couples Monte Carlo radiative transfer to the flow hydrodynamics. The code also self-consistently implements electron-positron pair production and annihilation. We simulate shocks with properties relevant for GRBs and study the steady-state solutions, which are accurate deep below the jet photosphere. The shock generates a power-law photon spectrum through the first-order Fermi mechanism, extending upwards from the typical upstream photon energy. Strong shocks (for which the downstream pressure is much larger than the upstream pressure) have rising $\nu F_\nu$ shock spectra. The spectrum extends up to $\epsilon_{max} \equiv E_{max}/m_e c^2 \sim v^2$ for non-relativistic shocks, where $m_e$ is the electron rest mass and $v$ is the relative speed between the upstream and downstream in units of the speed of light $c$. For mildly relativistic shocks the power law softens at $\epsilon \gtrsim 10^{-1}$ due to Klein-Nishina effects, and shocks with $v\gamma \simgt 1$, where $\gamma \equiv (1-v^2)^{-1/2}$, produce electron-positron pairs. As an example, a strong shock with $v\gamma = 3$ and a photon-to-proton ratio of $n_\gamma/n_p = 2 \times 10^5$ has a peak pair-to-proton ratio of $Z_\pm \approx 225$. The main effect of pairs in a steady-state shock is to decrease its spatial width by a factor of $\sim Z_\pm$. The post-shock spectrum thermalizes in the downstream. In absence of emission and absorption processes, kinetic equilibrium at temperature $\theta_d \equiv kT_d/m_e c^2 \approx \overline{\epsilon}_d/3$ is reached at an optical depth of $\tau \gg \theta_d^{-1}$ behind the shock, where $\overline{\epsilon}_d$ is the average downstream photon energy. We discuss the importance of these results for observations of emission from sub-photospheric shocks.
\end{abstract}




\section{Introduction}
\label{sec:introduction}

Shocks are ubiquitous in astrophysics. In most cases, the shocks are collisionless, mediated by collective plasma effects. The shocks dissipate incoming upstream kinetic energy, converting it to downstream plasma internal energy. A fraction of the dissipated energy is given to the electrons and can be promptly radiated away.

In general, the mechanism of shock dissipation depends on the upstream plasma conditions. In particular, shocks which occur in environments which are optically thick to scattering can be mediated by radiation. The most prominent examples of radiation mediated shocks (RMSs) are found in supernovae (SN) and long gamma-ray bursts (GRBs). They occur at the early stages of the stellar explosion, following the energy release inside the optically thick star. Furthermore, speed variations within GRB jets are expected to lead to mildly relativistic internal shocks. They can occur below the jet photosphere, where the plasma is still optically thick. Energy dissipation by sub-photospheric shocks have long been proposed as a mechanism for producing the prompt GRB emission (see e.g. \citealt{EicLev:2000, MesRee:2000, PeeMesRee:2006, RydPee:2009, Gia:2012}). \citet{LevBro:2008}, \citet{BroMikLev:2011} and \citet{Lev:2012} emphasized that sub-photospheric shocks are in fact mediated by radiation (as opposed to collective plasma effects). Recently, \citet{Bel:2017} (hereafter B17) performed time-dependent simulations demonstrating RMS formation in unmagnetized and magnetized flows.

RMSs differ qualitatively from collisionless shocks in a few important ways. First, the dissipated kinetic energy is directly transferred to the radiation through the first order Fermi mechanism, as opposed to being transferred to photons via electron internal energy. Second, a RMS is at least a few photon mean free paths wide, generally much larger than the Larmor radius of charged particles, preventing particle acceleration. Third, relativistic RMSs can heat photons to energies above the electron rest mass, leading to electron-positron pair production inside and around the shock. B17 estimated that $\sim 10^2$ pairs per ion should be created in GRB RMSs.

The photon spectrum within the RMS depends on the number of photons that share the dissipated kinetic energy. In general, the total number of photons downstream of the shock is the sum of photons advected from the upstream and the new photons generated by the shock itself. Two qualitatatively different RMSs can then be identified, depending on the dominant photon source. The upstream is defined as photon-poor if the newly produced photons dominate the shock photon number \citep{BroMikLev:2011}. This is the case for shocks propagating in a cold stellar envelope. Relativistic photon-poor RMSs were studied by \citet{KatBudWax:2010} and \citet{BudEtAl:2010}. The upstream is defined as photon-rich if the advection of radiation from the upstream dominates the photon number downstream of the shock. This is typically the case for RMSs inside the GRB jet \citep{Lev:2012}, as the jet consists of photon-rich plasma which originated close to the hot central engine. RMSs propagating into photon-rich upstreams (approximately) conserve photon number.

Non-relativistic photon-rich RMSs were studied by \citet{Weaver:1976}, \citet{BlaPay:1981}, \citet{Rif:1988} and \citet{Bec:1988}. \citet{LevBro:2008} studied the shock structure of mildly relativistic photon-rich RMSs, under the assumption of negligible pair production, by taking moments of the radiative transfer equation, accounting for the strong anisotropy of the photon field in the shock transition region which occurs at relativistic speeds.

The recent work of B17 used time-dependent simulations that couple the plasma hydrodynamics to Monte Carlo radiative transfer through scattering. B17 showed that in the presence of sufficiently strong magnetic fields, a collisionless ``subshock'' will form inside the RMS, located close to the RMS downstream. The subshock width is comparable to the ion Larmor radius, which is much smaller than the wide RMS structure. A fraction of the total shock energy is dissipated in the subshock, heating the electron (or pair) and proton components. The hot electrons quickly cool by inverse Compton scatterings and emission of synchrotron photons, until they reach the Compton temperature of the downstream radiation. Additionally, the presence of neutrons can further complicate the shock structure and dissipation profile through nuclear collisions on a length scale which is longer than the Thomson mean free path (\citealt{Bel:2010}, B17). In this work, we limit our considerations to RMSs in the limit of vanishing neutron component and magnetic fields. We will also assume that the photon number is conserved in the RMS, which is a good approximation when magnetic fields are sufficiently weak. The role of magnetic fields and the subshock for photon production will be discussed in a separate paper (Lundman \& Beloborodov, in preparation).

In this work we present a newly developed code for radiation hydrodynamics. Our code includes explicit treatment of $\gamma\gamma$ pair production, which for the first time allows for a fully self-consistent, time dependent flow solution, where significant amounts of pairs are expected. Furthermore, the code utilizes an exact Riemann solver, which accurately captures any (collisionless) subshocks that might develop (depending on the upstream conditions). We focus on shocks that occur deep below the photosphere, where the scattering time is much smaller than the jet expansion time. Such shocks quickly settle into a quasi-steady state, on a few scattering times. Deep sub-photospheric shocks are therefore essentially plane parallel. We let the shocks propagate until they settle into a steady-state, and then examine the shock solutions.


The paper is structured as follows. We derive the equations of special relativistic Lagrangian radiation hydrodynamics in Section~\ref{sec:equations} (and Appendix~\ref{app:derivation}). Our numerical hydrodynamics and Monte Carlo radiative transfer implementations are presented and discussed in Sections~\ref{sec:hydrodynamics} and \ref{sec:radiation}, respectively. We qualitatively discuss GRB RMSs in Section~\ref{sec:qualitative}, and numerically explore GRB shocks under four qualitatively different conditions in Section~\ref{sec:numerical}, confirming the main points of the previous section. Finally, the results are discussed in Section~\ref{sec:discussion}.



\section{Equations of radiation hydrodynamics}
\label{sec:equations}


Plasma and radiation will be treated using two distinct numerical methods. We first consider the plasma, which will be treated as a single fluid (this approximation is discussed in Section~\ref{sec:discussion}).


Conservation of energy and momentum is represented by the vanishing divergence of the stress-energy tensor; $T^{\alpha\beta}_{;\beta} = 0$. Separating the stress-energy tensor into matter (including electron-positron pairs) and radiation parts, $T^{\alpha\beta} = M^{\alpha\beta} + R^{\alpha\beta}$, we can write

\be
M^{\alpha\beta}_{;\beta} = G^\alpha,
\label{eq:energy_momentum_cons}
\ee

\noindent where $G^\alpha \equiv -R^{\alpha\beta}_{;\beta}$ is considered as an energy and momentum source term for the fluid equations. The stress-energy tensor of the fluid is $M^{\alpha\beta} = u^\alpha u^\beta (\rho + e + p) + p \eta^{\alpha\beta}$, where $u^\alpha = \Gamma(1, \beta^i)$ is the four-velocity, $\beta^i$ is the three-velocity, $\Gamma \equiv (1-\beta_i \beta^i)^{-1/2}$ is the Lorentz factor, $\rho$ is the total rest mass density (ions, electrons and positrons), $e$ is the internal energy density, $p$ is the pressure, $\eta^{\alpha\beta}$ is the Minkowski metric of signature $(-,+,+,+)$ and we use units for which the speed of light equals unity.

Conservation of proton number and the equation for the pair number density are given by

\be
(\rhop u^\alpha)_{;\alpha} = 0
\label{eq:mass_cons}
\ee

\noindent and

\be
(n_\pm u^\alpha)_{;\alpha} = \dot{n}_\pm
\label{eq:pair_equation}
\ee

\noindent respectively, where $\rhop$ is the proton rest mass density, $n_\pm$ is the pair (electron and positron) number density and $\dot{n}_\pm$ is the net rate of pair production/annihilation, all measured in the local rest frame of the fluid.

In Appendix~\ref{app:derivation}, we rewrite Equations~(\ref{eq:energy_momentum_cons})~-~(\ref{eq:pair_equation}) in the Lagrangian form and plane-parallel geometry. This gives


\be
(V_p)_{,t} - \beta_{,m} = 0,
\label{eq:V_p}
\ee

\be
(E_p)_{,t} + (p \beta)_{,m} = V_p G^0,
\label{eq:E_p}
\ee

\be
(S_p)_{,t} + p_{,m} = V_p G^1
\label{eq:S_p}
\ee

\noindent and

\be
(Z_\pm)_{,t} = m_p V_p \dot{n}_\pm.
\label{eq:Z_pm}
\ee

\noindent Here $m$ is the Lagrangian mass coordinate ($\partial_m = \Gamma\rhop \partial_x$) and $\partial_t$ is the Lagrangian time derivative ($\partial_t \rightarrow \partial_t - \beta\partial_x$). The new variables $V_p$, $E_p$ and $S_p$ are the lab frame volume, energy and momentum per proton rest mass respectively, defined as $V_p = (\Gamma\rhop)^{-1}$, $E_p = V_p (\Gamma^2(\rho + e + p) - p)$ and $S_p = V_p \beta\Gamma^2(\rho + e + p)$. The pair loading factor is $Z_\pm = n_\pm/n_p$, and $m_p$ is the proton rest mass. The right hand sides of Equations~(\ref{eq:E_p}) and (\ref{eq:S_p}) are the rates of energy and momentum gain per proton, measured in units of the proton rest mass.


\section{Hydrodynamics implementation}
\label{sec:hydrodynamics}

The hydrodynamical Equations~(\ref{eq:V_p})~-~(\ref{eq:S_p}) are solved numerically using a standard Lagrangian scheme with an exact Riemann solver (e.g. \citealt{DaiMoc:2000}). In short, we discretize the equations (including Equation~(\ref{eq:Z_pm})) into fluid elements of a given proton mass, using finite differences on a mass grid. The time evolution of each fluid element is then integrated in the following way. The Piecewise Parabolic Method (PPM) \citep{ColWoo:1984}, is used to find ``left and right states'' of the fluid at each grid interface (and time step). The left and right states are used to solve a Riemann problem, in order to find the spatial (mass) derivative approximations.

A more detailed description of the hydrodynamics implementation is provided in the following subsections. The source terms ($G^\alpha$ and $\dot{n}_\pm$) are obtained from the Monte Carlo radiative transfer, as described in Section~\ref{sec:radiation}.

\subsection{Discretization of the hydrodynamical equations}

The Lagrangian grid is defined by the value of the mass coordinate at the interfaces between grid cells, and each grid cell represent a ``fluid element''. The fluid element mass equals the difference between the mass coordinate values at the cell boundaries: $\Delta m_j = m_{j+1/2} - m_{j-1/2}$, where $j$ labels a cell, $j \pm 1/2$ labels the right and left cell boundaries and $\Delta m_j$ is the proton mass contained in the cell.

Each cell contains (mass averaged) values of $V_p$, $S_p$ and $E_p$, or equivalently $\beta$, $\rho$ and $p$. The PPM method and a Riemann solver (as described below) are used to find time averaged values of the pressure, $\bar{p}$ and velocity, $\bar{\beta}$ at each cell boundary, given $\beta$, $\rho$ and $p$ inside each cell. The time averaged values at the cell boundaries are then used for approximating the spatial derivatives ($\partial_m$) and updating the properties of each fluid element.

Each mass boundary $m_{j+1/2}$ is associated with a boundary in the spatial coordinate, $x_{j+1/2}$, representing the fluid element boundary location in space (at a given time). The motion of each boundary during the time step $\Delta t$ is discretized by a simple finite difference in time. The spatial grid at time step $n+1$ is then given by the equation of motion of the boundary and the grid at the previous time step as,

\be
x_{j+1/2}^{n+1} \approx x_{j+1/2}^n + \bar{\beta}_{j+1/2}^n \Delta t,
\label{eq:x_finite_difference}
\ee

\noindent and the updated volume per proton is given by

\be
(V_p)_j^{n+1} \approx \frac{x_{j+1/2}^{n+1} - x_{j-1/2}^{n+1}}{\Delta m_j},
\ee

\noindent implicitly solving Equation~(\ref{eq:V_p}). Straight-forward discretization of the hydrodynamical equations for energy and momentum (Equations~(\ref{eq:E_p}) and (\ref{eq:S_p})) gives

\begin{eqnarray}
(E_p)_j^{n+1} & \approx & (E_p)_j^n + (V_p)_j^n (G^0)_j^n \Delta t \nonumber\\
              &         & - \frac{\Delta t}{\Delta m_j} \left[\bar{p}_{j+1/2}^n \bar{\beta}_{j+1/2}^n - \bar{p}_{j-1/2}^n \bar{\beta}_{j-1/2}^n\right]
\end{eqnarray}

\noindent and

\begin{eqnarray}
(S_p)_j^{n+1} & \approx & (S_p)_j^n + (V_p)_j^n (G^1)_j^n \Delta t \nonumber\\
              &         & - \frac{\Delta t}{\Delta m_j} \left[\bar{p}_{j+1/2}^n - \bar{p}_{j-1/2}^n\right].
\label{eq:S_finite_difference}
\end{eqnarray}

\noindent Similarly, the pair loading equation is discretized as

\begin{eqnarray}
(Z_\pm)_j^{n+1} & \approx & (Z_\pm)_j^n \nonumber\\
                &         & + m_p (V_p)_j^n (\dot{n}_\pm)_j^n \Delta t.
\end{eqnarray}

\subsection{Variable reconstruction}

The variables $\beta$, $\rho$ and $p$ must be reconstructed numerically from $V_p$, $E_p$ and $S_p$ in each grid cell and for each time step. Following \citet{DaiMoc:2000}, we numerically solve the equation

\be
\hs^2 + (\gamma_{ad}-1) \hs - \gamma_{ad} E_m (S_m^2 + \hs^2)^{1/2} + \gamma_{ad} S_m^2 = 0
\label{eq:Taub}
\ee

\noindent for the specific enthalpy, $\hs \equiv h/\rho = 1 + (e+p)/\rho$, $\gamma_{ad}$ is the adiabatic index of the fluid and $E_m = E_p /(1 + Z_\pm m_e/m_p)$ is the energy per unit total mass (and similar definitions hold for $S_m$ and $V_m$). We use the Newton-Raphson method to solve Equation~(\ref{eq:Taub}). After $\hs$ is found, $\beta$, $\rho$ and $p$ are computed as $\beta = S_m (S_m^2 + \hs^2)^{-1/2}$, $\rho = (\Gamma V_m)^{-1}$ and $p = \rho (\hs-1) (\gamma_{ad}-1) / \gamma_{ad}$.

\subsection{Finding left and right states using the PPM method}

The PPM method is an extension of Godunov's method, with the advantage of being second order accurate in time and third order in space. Parabolic (quadratic) polynomials are fit to each of the hydrodynamic grid quantities $\beta$, $\rho$ and $p$ at a given time step. The parabolic fits provide continous representations of the hydrodynamic quantities.

Only a fraction of the fluid in each grid cell can affect the conditions at the cell boundary during a time step (assuming that the time step is short enough for the computation to converge). The distance into each cell from which information can reach the cell boundary is found, and the time and mass averaged values of $\beta$, $\rho$ and $p$ are computed using the continous polynomials at each side of the boundary. The averaged values at each side of the boundary define the left and right fluid states which are needed for solving the Riemann problem at the boundary. We refer to \citet{ColWoo:1984} for a detailed discussion of the parabolic fits and the averaging process.

\subsection{Solving the Riemann problem}

A Riemann solver is designed to numerically compute the pressure and speed of the intermediate region which develops in interaction between two initially separated fluid states. The intermediate region includes the contact discontinuity, which separates the fluid that was originally contained in the left and right states. Since we are solving Lagrangian equations, the contact discontinuity of the Riemann problem directly corresponds to the boundary between two grid cells.

The Riemann problem admits three qualitatively different solution patterns; two shocks, one shock and one rarefaction wave, or two rarefaction waves can be launched. We use the exact, special relativistic Riemann solver developed by \citet{RezZan:2001}. It has the advantage of determining the solution pattern directly from the initial conditions. This makes the numerical implementation simpler, as the functional form of the solution is known before attempting numerical convergence. The Riemann solver gives $\bar{p}$ and $\bar{\beta}$ at each interface, which are then used to solve Equations~(\ref{eq:x_finite_difference})~-~(\ref{eq:S_finite_difference}), updating $V_p$, $E_p$ and $S_p$, and completing the hydrodynamical time step.

\section{Radiation implementation}
\label{sec:radiation}

The radiative transfer is performed using the Monte Carlo method. The radiation is described by discrete Monte Carlo photons, or photon ``packets''. A photon packet is defined by its spatial location ($x$), direction relative to the spatial axis ($\mu = \cos\theta$), energy ($\epsilon = E/m_e c^2$) and its weight ($w$). The photon packet weight gives the number of real photons (with assumed identical properties) represented by the packet (or more precisely, photons per unit area, due to the assumed planar symmetry of the problem). The weight is initially computed as

\be
\Delta N_\gamma w = \Gamma n_\gamma \Delta x,
\label{eq:weight}
\ee

\noindent where $\Delta N_\gamma$ is the chosen number of Monte Carlo photons within the spatial bin of width $\Delta x$, and $n_\gamma$ is the photon number density, as set by the initial conditions. An array of all photon packets is kept in memory. The photons propagate through the hydrodynamical Lagrangian grid, with different radiative processes contributing to the opacity. Photon packets are added to or removed from the array as soon as they are emitted or absorbed by the plasma, with the corresponding energy and momentum differences subtracted or added to $G^\alpha$ at the relevant grid location as to conserve energy and momentum. In the text below we will refer to photon packets simply as photons for brevity.

For this work we consider only the radiative processes of scattering and $\gamma\gamma$-absorption. However, additional interactions can be added if needed, as no specific number of radiative processes is assumed in the code description below.

\subsection{Propagation algorithm}

The code picks a time step $\Delta t$ and then iterates over each photon in the array. The selected photon propagates and interacts with the plasma until it is either absorbed or it has propagated for time $\Delta t$. The propagation algorithm is logically separated into ``events''. An event is here defined as either an interaction (scattering or absorption) or the crossing of a grid cell boundary into a neighboring mass bin. A typical propagation step consists of zero to several events, and most events are grid crossings.

Below is a more detailed description of the propagation algorithm. First, the mass bin where the photon is located is found by bisection (information regarding the photon location during the previous time step can be used here). The code computes local mean free paths in the photon propagation direction for all relevant radiative processes, using the plasma properties of the current spatial bin. (The computation of the mean free path for $\gamma\gamma$-annihilation is described in the next subsection.) The total mean free path $\lambda$ is computed by adding the absorption coefficients for each process (e.g. $\lambda^{-1} = \sum_i \lambda_i^{-1}$, where $i$ labels each process). A lab frame propagation distance $l$ is drawn from the exponential distribution as $l = -\lambda \ln(u)$, where $u$ is a random number uniformly distributed between zero and one. If time $\Delta t$ has passed before an event occurs, the photon is simply propagated for the remaining time. Otherwise, the code moves the photon to the event location, updates the photon propagation time and performs the event. The event type is determined by whatever happens first, either propagating a distance $l$ or crossing a boundary. The crossing simply consists of moving the photon to the current boundary location, updating the current bin location. The boundaries are assumed to move during photon propagation, at a speed equal to the average speed of the neighboring bins. This gives better accuracy in flows with large bulk motion. As soon as an event has occurred, the algorithm computes new mean free paths $\lambda_i$ and draws a new $l$. The timestep is completed for the photon when its propagation time reaches $\Delta t$ or there is an absorption event.

At the end of the photon free path $l$, the code must determine which radiative process has occurred. The probability for process $i$ not to occur over distance $l$ is given by

\be
F_i(l;\lambda_i) = \int\limits_l^\infty f_i(l^\prime;\lambda_i) \D l^\prime,
\ee

\noindent where $f_i(l^\prime;\lambda_i)$ is the probability density for distance $l^\prime$ and process $i$. The probability for process $i$ to occur in the interval $(l, l+\delta l)$, and at the same time be the first process that occurs, is

\be
\delta p_i = f_i \delta l \prod\limits_{j \neq i} F_j,
\ee

\noindent where the probabilities are simply multiplied, since they are independent. The probability density distribution for each process is the exponential distribution, for which $f_i(l;\lambda_i) = \lambda_i^{-1} \exp(-l/\lambda_i) = \lambda_i^{-1} F_i$, and therefore $\delta p_i = \lambda_i^{-1} \delta l \prod F_j$, where the product now runs over all $j$.

The probability that the $i$-th interaction process occurs at $l$ is then found as the ratio $P_i = \delta p_i/\sum \delta p_i$, where the sum is taken over all processes $i$. We then arrive at

\be
P_i = \frac{\lambda_i^{-1}}{\sum\limits_j \lambda_j^{-1}},
\ee

\noindent which is independent of the value of $l$.

\subsection{Scattering}

The code uses the full Klein-Nishina cross-section for computing the scattering mean free path (as a function of photon energy and direction) and scattering angles. The gas temperature is determined by the hydrodynamic internal energy of the gas, and the electron (and positron) distribution is assumed to be Maxwellian. Relaxation to kinetic equilibrium between photons and electrons through scatterings have been tested extensively, with an initially non-thermal photon spectrum relaxing to the Wien spectrum (as expected in absence of stimulated scattering), while conserving energy.

\subsection{Pair production and annihilation}

The mean free path to $\gamma\gamma$-annihilation is not a function of the local plasma properties, but requires knowledge on the local radiation intensity. We define a grid in the photon energy ($\epsilon^\prime$) and direction ($\mu^\prime$), both measured in the fluid rest frame (the comoving frame). At each time step and grid cell, the comoving intensity is computed on the two-dimensional $(\epsilon^\prime, \mu^\prime)$ grid by collecting the Monte Carlo photons. The mean free path can be considered a function of $\epsilon^\prime$, $\mu^\prime$, and the photon location $x$, $\lambda^\prime_{\gamma\gamma}(x, \epsilon^\prime, \mu^\prime)$. Computation of $\lambda^\prime_{\gamma\gamma}$ (measured in the fluid comoving frame) for one particular set of location, energy and direction involves integration of the target photon number intensity, $\mathcal{I}^\prime_{\nu^\prime}$, over the target energy and direction,

\be
(\lambda_{\gamma\gamma}^\prime)^{-1} = \int (1-\tilde{\mu}^\prime) \sigma_{\gamma\gamma} \mathcal{I}^\prime_{\nu^\prime} \D\Omega^\prime \D\nu^\prime,
\ee

\noindent where $\sigma_{\gamma\gamma}$ is the center-of-momentum frame cross-section and $\tilde{\mu}^\prime$ is the cosine of the angle between the primary and target photon directions. The mean free path $\lambda^\prime_{\gamma\gamma}(x, \epsilon^\prime, \mu^\prime)$ is tabulated on the grid of $x$, $\epsilon^\prime$, $\mu^\prime$ at each time step, before the propagation of photons is initiated.

The comoving mean free path for a photon is found by bisection in the grid of $\epsilon^\prime$ and $\mu^\prime$. The lab frame mean free path is then obtained by a Doppler boost, $\lambda_{\gamma\gamma}(\epsilon, \mu) = \lambda^\prime_{\gamma\gamma}(\epsilon^\prime, \mu^\prime) / [\Gamma(1-\beta\mu)]$.

The pair production source term can be written as $\dot{n}_\pm = \dot{n}_\mathrm{\pm,prod} - \dot{n}_\mathrm{\pm,ann}$. Each fluid element has a proton mass of $\Delta m$, which corresponds to $\Delta m/m_p$ protons (per area), and a Monte Carlo photon packet corresponds to $w$ photons (per area). After a photon-photon interaction the photon is absorbed, adding an equal number of pairs (per photon packet) to the plasma,

\be
\frac{\delta n_\mathrm{\pm,prod}}{n_p} = \frac{m_p w}{\Delta m}.
\ee

\noindent Requiring that a single annihilation event changes $Z_\pm$ only slightly, i.e. $\delta Z_\pm = \delta n_\mathrm{\pm,prod}/n_p \ll 1$, we find a lower limit on the number of photon packets per bin; $\Delta N_\gamma \gg n_\gamma/n_p$, where Equation (\ref{eq:weight}) was used.

At each time step the code computes the number of real photons which are emitted due to pair annihilation, based on the number of annihilated pairs, $\delta n_\pm = \dot{n}_\mathrm{\pm,ann} \Delta t$, where $\dot{n}_\mathrm{\pm,ann} = (3/4) \sigmat n_- n_+$ and $n_-$ ($n_+$) is the number density of electrons (positrons). Using the relations $n_\pm = n_- + n_+$ and $n_p = n_- - n_+$, we have

\be
\frac{\dot{n}_\mathrm{\pm,ann}}{n_p} = \frac{3}{16} \sigmat n_p (Z_\pm^2 - 1).
\label{eq:pair_annihilation}
\ee

\subsection{Photon boundary conditions}

The hydrodynamic code is a Lagrangian code, which tracks the motion of individual fluid elements. The hydrodynamic boundary conditions used for the simulations presented in this paper consists of one reflective (lab frame) wall and one ``comoving wall''; that is, two walls are assumed to exist on each side of the grid, but one of the walls is moving toward the other. The reflective wall simply reflects the lab frame grid speed, and also all photons which propagate into it. The comoving wall is moving with the initial speed of the outermost fluid element. The corresponding photon boundary condition can be stated as ``zero comoving frame flux'' through the grid outer boundaries. This amounts to reflecting photons in the local comoving frame as they try to escape the grid edge (each reflected photon can also be viewed as a new photon with identical properties but opposite direction, as viewed in the comoving frame of the boundary, while the old photon was allowed to leave the grid).

\subsection{Electron cooling time and length scales}

Consider a plasma with a certain number of photons per proton, $n_\gamma/n_p$, a pair loading factor of $Z_\pm$ and a typical photon energy $\epsilon = E/m_e c^2$. The scattering time for the photons is $t^{-1}_{sc} \approx Z_\pm n_p \sigmat$; the photon time step must resolve (at least) the scattering time, as this is the characteristic timescale for the RMS.

Consider now a collisionless subshock located close to the immediate RMS downstream, where electrons are heated up to a Lorentz factor $\gamma_e$. The corresponding electron cooling time is $t^{-1}_{cool} \approx (4/3) \xiKN \gamma_e \epsilon n_\gamma \sigmat$, where $\xiKN \approx (1 + 4\gamma_e\epsilon)^{-3/2}$ describes the decrease of inverse Compton cooling efficiency due to Klein-Nishina effects \citep{ModEtAl:2005}, so that

\be
\frac{t_{cool}}{t_{sc}} \approx 3 Z_\pm \frac{n_p}{n_\gamma} \frac{(1+4\gamma_e\epsilon)^{3/2}}{4\gamma_e\epsilon}
\ee

\noindent The number of photons per proton is in the range of $10^4 - 10^6$ for typical GRB conditions, and the electron cooling time is therefore typically several orders of magnitude shorter than the photon scattering timescale (depending on $\epsilon$). The photon time step must resolve the cooling timescale, as photons may otherwise ``break the energy budget'' by interacting with high-energy electrons for too long, consuming all internal fluid energy before the fluid can react by lowering its temperature. The corresponding length scale, $l_{cool}$, must also be spatially resolved to capture accurate electron temperatures behind a subshock.

A mildly relativistic RMS is a few scattering free paths wide, and its development and dynamics are therefore related to the photon scattering time. Resolving the electron cooling time may then be computationally challenging. However, a numerical ``trick'' can be used for RMS without a significant subshock. In this case, the fact that $n_\gamma \gg Z_\pm n_p$ implies that the electrons/positrons are locked to the Compton temperature everywhere, including within the RMS. One can then add a ``fake heat'' reservoir to the plasma initial conditions: simply multiply the internal energy by a constant factor which is larger than unity, and divide by the same factor when computing the fluid temperature as seen by the photons. This trick artificially increases the internal energy budget (which must still be much smaller than the photon energy budget), and therefore also increases the plasma cooling time, permitting larger time steps. The upper limit on the amount of fake heat is set by the condition that the fluid is still locally locked to the Compton temperature.

\subsection{Code parallelization}

Monte Carlo codes are easily parallelizable. Our implementation initiates a given number of photon packets on each CPU core. The hydrodynamics computations (which takes much less CPU time compared with photon propagation) are performed only on the master core. At each time step, the master distributes the hydrodynamic grid to the worker cores. The workers propagate their photons in the grid and compute all hydrodynamical sources ($G^\alpha$ and $\dot{n}_\pm$). The master then collects the sources from all workers for the next hydrodynamics step. Similarly, the comoving radiation intensity (which is used for the pair production algorithm) is computed locally by each worker, and then collected by the master. The computation of the $\gamma\gamma$-annihilation mean free path is distributed over all workers, as this computation is fairly expensive.

\section{RMS in GRB jets}
\label{sec:qualitative}

Shocks that occur deep below the GRB jet photosphere have a characteristic width which is much smaller than any macroscopic flow length scale. Such shocks are therefore essentially plane parallel and in quasi-steady state. In this section we discuss the properties of steady plane-parallel RMS.


\subsection{The shock spectrum}

Given upstream values of $w \equiv (e_\gamma + p_\gamma)/\rho$ and $v\gamma$, where $v$ is the upstream speed relative to the downstream (in units of the speed of light) and $\gamma$ is the corresponding Lorentz factor, one can solve for the downstream $w$ and shock compression ratio (B17). Besides these thermodynamic parameters, the photon spectrum in the shock transition region depends on the photon-to-proton ratio, $n_\gamma/n_p$, of the upstream material \citep{Lev:2012}. A RMS converts the incoming kinetic proton energy to radiation energy; each proton shares its energy with $n_\gamma/n_p$ photons. The conversion is complete in the immediate downstream, so that the average photon energy must equal (in the limit of a cold upstream, $w_u \ll 1$)

\begin{eqnarray}
\bar{\epsilon}_d &       = & (\gamma-1)\frac{m_p}{m_e}\frac{n_p}{n_\gamma} \nonumber\\
                 & \approx & 1.8 \times 10^{-2} \, (\gamma-1) \left(\frac{n_\gamma/n_p}{10^5}\right)^{-1}.
\label{eq:epsilon_d}
\end{eqnarray}

Photons gain energy inside the shock by scattering repeatedly within the converging fluid flow (i.e. the first order Fermi mechanism). In order for photons to gain energy in the shock, so that they can mediate it, the RMS must structure itself so that the ``shock $y$-parameter'' is of order unity; $(\Delta\epsilon/\epsilon) N_{sc} \sim 1$, where $\Delta\epsilon/\epsilon$ is the fractional energy gain per scattering and $N_{sc}$ is the typical number of scatterings for a photon which diffuses through the shock structure. The optical depth of a non-relativistic RMS transition is $\tau_{sh} \sim v^{-1}$, and the number of scatterings is $N_{sc} \sim \tau_{sc}^2$. Thus, the typical fractional energy gain per scattering in a non-relativistic RMS is

\be
\frac{\Delta\epsilon}{\epsilon} \sim v^2.
\label{eq:gain}
\ee

Since the photons greatly outnumber the electrons (and positrons) in a GRB RMS, the electrons carry essentially no heat capacity and are locally locked to the Compton temperature $\theta_C \equiv kT_C/m_e c^2$ (defined as the electron temperature for which there is no net energy exchange between electrons and photons through scatterings). The Compton temperature close to the immediate downstream is roughly $\theta_C \sim \bar{\epsilon}_d$ (within a factor of a few, depending on the spectral shape). Photons of energy $\epsilon \ll 4\theta_C$ will not only gain energy by scattering in the speed gradient, but also experience thermal Comptonization. The average thermal energy gain per scattering is $\Delta\epsilon_\mathrm{th}/\epsilon \approx 4\theta_C \approx 4 (v^2/2) (m_p n_p)/(m_e n_\gamma)$, where Equation~(\ref{eq:epsilon_d}) was used. The thermal energy gain is much smaller than the energy gain due to scattering inside the converging flow as long as

\be
\frac{n_\gamma}{n_p} \gg \frac{m_p}{m_e},
\ee

\noindent which is easily satisfied for GRBs which have $n_\gamma/n_p \sim 10^5$. Thus, thermal Comptonization can be neglected inside the shock.

A fraction of photons scatter back toward the upstream, and spend longer time in the RMS, gaining more energy. Since both the relative energy gain, $\Delta\epsilon/\epsilon$, and the scattering cross-section are independent of photon energy (for $\epsilon \ll 1$), and thermal Comptonization can be neglected, the problem now lacks an energy scale, and the photon spectrum will form a power law which extends upward from the typical upstream photon energy $\epsilon_u$ (as shown by \citealt{BlaPay:1981} in the non-relativistic limit).


Mildly relativistic (and faster) shocks are similar; the shock $y$-parameter has to be of order unity. Just as for non-relativistic shocks, there is a significant chance for a photon that just exited the shock in the downstream to scatter back into the shock and continue the energy gain. The resulting spectrum is again a power law.

The power law can be at most flat in $\nu F_\nu$ (as found by \citealt{BlaPay:1981} in the non-relativistic limit). A flat power law would imply a logarithmic divergence of radiation energy, however an upper photon energy naturally exists due to electron recoil (and also pair production at energies above $m_e c^2$). Considering only the recoil, the energy after scattering in a direction $\mu_{sc} = \cos\theta_{sc}$ is $\epsilon_1/\epsilon = 1/(1 + \epsilon(1-\mu_{sc})) \approx 1/(1 + \epsilon) \approx 1 - \epsilon$ (where we substituted the average $\mu_{sc} = 0$ for scattering of photons with $\epsilon\ll 1$). Therefore the typical energy loss due to recoil is


\be
\frac{\Delta\epsilon_\mathrm{recoil}}{\epsilon} \approx -\epsilon.
\label{eq:recoil}
\ee

\noindent For non-relativistic shocks, the energy gains and losses (Equations~(\ref{eq:gain}) and (\ref{eq:recoil})) balance at photon energies

\be
\epsilon_{max} \sim v^2.
\ee

\noindent We see that $\epsilon_{max} \sim 1$ for mildly relativistic shocks which have $v\gamma \sim 1$. Pair production is therefore expected to become relevant for shocks with $v\gamma \gtrsim 1$.

The shock structure will self regulate into a shape that produces a photon spectrum with an average photon energy of $\bar{\epsilon}_d$ in the immediate downstream. The exact shape of the high-energy spectrum is challenging to predict for mildly relativistic (and faster) shocks without resorting to accurate simulations because of the coupled dynamics of the system and the non-trivial radiative transfer at energies close to and above the electron rest mass. The decrease of the Klein-Nishina cross-section with energy causes the mean free path to increase, so that photons can more easily propagate across the full width of the shock in a single free path. On the other hand, the longer mean free path can take the photons far downstream where they may become trapped after losing energy in scattering (which increases their scattering cross-section). Furthermore, high-energy photons scatter preferentially along their own forward direction, so that it becomes unlikely for the photon to turn around and scatter back toward the shock. If the photon manages to scatter at a large angle so that it may catch up with the shock, it will lose a significant fraction of its energy to electron recoil, and thus have its mean free path become shorter, decreasing the probability to reach the shock. We therefore expect the power law spectrum to soften at energies around $\epsilon \gtrsim 10^{-1}$ (assuming that the shock is capable of producing photons of such large energies).

The power law index is related to the ratio of the average photon energy in the downstream, $\bar{\epsilon}_d$, to the characteristic  thermal photon energy in the upstream $\epsilon_u$. Consider for instance a flat $\nu F_\nu$ spectrum inside the shock and suppose for simplicity that it extends to $\epsilon \sim 1$; then $\bar{\epsilon}_d/\epsilon_u \sim \ln(\epsilon_u^{-1})$, or $\bar{\epsilon}_d \sim 10 \epsilon_u$ for $\epsilon_u \sim 10^{-4}$. If larger values of $\bar{\epsilon}_d/\epsilon_u$ are demanded by the shock, then the spectrum must be rising in $\nu F_\nu$. If the spectrum is instead flat in photon number, such that $\nu F_\nu \propto \epsilon$, then the average energy is $\bar{\epsilon}_d \approx 1/\ln(\epsilon_u^{-1}) \sim 10^{-1}$ for $\epsilon_u \sim 10^{-4}$. Thus, strong shocks with $\bar{\epsilon}_d \gg \epsilon_u$ are expected to have (approximately) power law spectra inside the shock which range from flat in energy per decade to flat in photon number per decade, and the spectra are expected to deviate from the power law at $\epsilon \sim 10^{-1}$ in mildly relativistic shocks.

If the shock spectrum extends to an energy $\epsilon_{max} > 1$, then the high-energy photons tend to pair produce on photons of energy $\epsilon \approx 1/\epsilon_{max}$. The spectrum at $\epsilon \sim \epsilon_{max}$ can then be softened by $\gamma\gamma$-absorption. If the photon number spectrum is almost flat at $\epsilon \sim 1$, so that roughly equal number of photons exists at $\epsilon_{max}$ and $1/\epsilon_{max}$, then the spectrum in the whole energy range $1/\epsilon_{max} \lesssim \epsilon \lesssim \epsilon_{max}$ is softened.

\subsection{The peak pair multiplicity}

An upper limit on the number of pairs that a shock can sustain is set by assuming that all of the dissipated shock energy is converted into pair rest mass; $Z_\pm m_e = (\gamma-1) m_p$, so that $Z_\pm \lesssim 10^3$ for a mildly relativistic shock. However, the bulk of the proton energy is not channeled into photons with $\epsilon > 1$ as long as $\bar{\epsilon}_d \ll 1$, and so $Z_\pm \ll (\gamma-1)m_p/m_e$ is expected.

As the shock is initially formed, the number density of photons with energy $\epsilon \sim 1$ is small and pair creation is inefficient. The shock keeps building up its high-energy photon component until the photon loss rate at large energies becomes comparable to the rate of high-energy photon production. The rate of production is related to the scattering timescale, and the loss rate is related to the $\gamma\gamma$-annihilation timescale (and also advection into the downstream). With similar rates at $\epsilon \simgt 1$, the mean free paths to scattering and $\gamma\gamma$-annihilation are also similar, $\lambda_{\gamma\gamma} \sim \lambda$ where $\lambda_{\gamma\gamma} \sim (n_\mathrm{HE} \sigma_\mathrm{T}/5)^{-1}$ is the mean free path to annihilation and $n_\mathrm{HE} = f_\mathrm{HE} n_\gamma$ is the density of photons with energies $\epsilon \sim 1$. The scattering mean free path is increased by roughly a factor of $5$ at $\epsilon \sim 1$ due to Klein-Nishina effects, so that $\lambda \sim (Z_\pm n_p \sigma_\mathrm{T} / 5)^{-1}$. We then find that $Z_\pm \sim f_\mathrm{HE} (n_\gamma/n_p)$.

As mentioned above, the radiative transfer at $\epsilon \sim 1$ is complicated due to Klein-Nishina effects and is therefore best evaluated numerically. For example, one of the simulations presented in the next section (``Faster shock into cold upstream'') has $n_\gamma/n_p = 2 \times 10^5$ and develops a spectrum reaching up to $\epsilon_{max} \sim 3$. We found $n_\gamma(\epsilon > 1/3) \sim 10^{-3} n_\gamma$ inside the shock, so that the estimated pair multiplicity is $Z_\pm \sim 200$, close to the value of $Z_\pm \approx 225$ found in the simulation. Our results confirm the estimate for $Z_\pm$ in B17.

\subsection{The upstream photon precursor}

A fraction of the shock photons will leak ahead of the shock into the upstream, and pre-heat the upstream plasma through scatterings. Scattered photons isotropize and propagate with the upstream plasma, so that even the isotropic component of the photon spectrum inside an upstream fluid element becomes increasingly non-thermal as the shock is approaching. The strength of the photon precursor naturally weakens with distance into the upstream as the photon beam is attenuated. Photons with energies $\epsilon \gtrsim 10^{-1}$ have longer mean free paths due to the energy dependence of the Klein-Nishina cross-section, and therefore propagate further than low energy photons, hardening the upstream spectrum somewhat with distance from the shock. Neglecting the fact that the scattering cross-section is energy dependent, the intensity of the photon precursor is proportional to $\exp(-\tau)$ where $\tau$ is the total Thomson optical depth (including pairs) as measured from the shock into the upstream.

If the shock contains photons of energies greater than the electron rest mass, then the precursor will also sprinkle pairs in the upstream, ahead of the shock. The high-energy photons close to the shock can easily collide and convert to pairs. At larger distances into the upstream the photon precursor quickly becomes collimated in the forward direction, so that photons of energy $\epsilon \gtrsim 1$ can not efficiently pair produce on each other. On the other hand, the photons are free to produce pairs as soon as one of the photons scatters, so that the angle between the photons increases. The rate of pair production in the upstream is then tied to the scattering rate.

We can estimate the pair loading dependence on the distance into the upstream in the following way. Consider a steady-state shock. The pair loading equation (Equation~(\ref{eq:Z_pm})) can be written as $(Z_\pm)_{,t} = \dot{n}_\pm/\gamma n_p$. The equation of motion for a fluid element which is advected from the upstream toward the shock is $x_{,t} = - v_u$, where the $x$ coordinate is measured from the shock toward the upstream (in the shock frame) and $v_u > 0$ is the upstream speed relative to the shock. The pair loading equation for a fluid element can then be written as

\be
\left(Z_\pm\right)_{,x} = - \frac{\dot{n}_\pm}{\gamma_u v_u n_p}.
\ee

\noindent If we assume that all scattered high-energy photons are immediately converted to pairs, and that pair annihilation is negligible, then the net pair production rate is $\dot{n}_\pm \sim 2 (\sigma_\mathrm{T}/5) Z_\pm n_p n_\mathrm{HE}$ (with two pairs created for each scattering, and the scattering mean free path approximately five times the Thomson mean free path), where $n_\mathrm{HE} \propto \exp(-\tau)$ is the number density of high-energy photons from the pre-cursor, and $\tau \approx \gamma_u (1+v_u) \int Z_\pm n_p \sigma_\mathrm{T} \mathrm{d}x$ is the optical depth into the upstream as measured from the shock. Changing the variable from $x$ to $\tau$, we find $(Z_\pm)_{,\tau} \propto -\exp(-\tau)$, and integrating this equation from far in the upstream toward the shock, we find $Z_\pm \propto \exp(-\tau)$ (for $Z_\pm \gg 1$), which agrees well with the simulation results shown in the next section.

\subsection{Downstream spectrum ``thermalization''}

The shock spectrum is highly non-thermal. The photons ``thermalize'' (or rather, approach kinetic equilibrium with the electrons) in the downstream by re-distributing their energy through scatterings. The thermal Compton $y$-parameter is $y_{th} = 4\theta_C N_{sc} \sim \bar{\epsilon}_d N_{sc}$ where $N_{sc}$ is the number of scatterings. Low energy photons can significantly increase their energy when $y_{th} \gtrsim 1$, or $N_{sc} \gtrsim 1/\bar{\epsilon}_d$. The number of scatterings $\delta N_{sc}$ performed in time $\delta t$ is $\delta N_{sc} \approx \delta t/t_{sc}$, where $t_{sc} = \lambda_{sc}$ is the scattering time, $\delta t \approx \delta l/v_{sh} \approx 3 \delta l$, $\delta l$ is the distance behind the shock and $v_{sh} \approx 1/3$ is the shock speed in the downstream frame of a relativistic shock. The number of scatterings is then related to the downstream optical depth, $\delta \tau = \delta l/\lambda_{sc}$, as measured from the shock and into the downstream as $\delta N_{sc} \approx 3 \delta \tau$. Integrating the number of scatterings over the distance behind the shock, we then find the thermalization optical depth,

\be
\tau_{th} \sim \frac{1}{3 \bar{\epsilon}_d}.
\ee

\noindent If $\bar{\epsilon}_d \ll 1$, the low energy power law spectrum is modified far away from the shock.

On the other hand, the high-energy spectrum at $\epsilon > \bar{\epsilon}_d$ is affected more quickly by downscattering, as only $N_{sc} \approx (\Delta\epsilon/\epsilon)^{-1} \sim 1/\epsilon$ scatterings are needed to modify the photon energy. Thus, the spectrum at the highest energies is expected to progressively soften into the downstream, starting already inside the shock at $\epsilon \gtrsim 10^{-1}$.


The pairs that were produced inside the shock annihilate as they propagate into the downstream. In steady-state the pair multiplicity satisfies $v_{sh} (Z_\pm)_{,x} = \dot{n}_\pm/\gamma n_p$ in the shock frame, where $x$ is the distance as measured from the shock into the downstream. Pair production quickly ceases in the downstream, so that only pair annihilation is important; $\dot{n}_\pm \approx -\dot{n}_{ann}$. From Equation~(\ref{eq:pair_annihilation}) we find (for $Z_\pm \gg 1$)

\be
\delta Z_\pm \approx - \frac{Z_\pm^2 \sigma_\mathrm{T} n_p}{5 v_d \gamma_d} \delta x.
\ee

\noindent The number of scatterings experienced by a downstream photon in time $\delta t$ is $\delta N_{sc} \approx \delta t/\gamma t_{sc} = \delta x / v_{sh}\gamma \lambda_{sc}$, where $\lambda_{sc} = 1/Z_\pm \sigma_\mathrm{T} n_p$. We then find $\delta \ln Z_\pm \approx - \frac{1}{5} \delta N_{sc}$, with the solution

\be
N_{sc} \approx 5 \ln\left(\frac{Z_{max}}{Z_\pm}\right).
\ee

\noindent A typical photon which has passed through a shock with $Z_{max} \sim 200$ will have scattered $N_{sc} \sim 25$ times before the pairs are annihilated.

\subsection{The internal GRB RMS parameter space}

The total lab frame energy per proton rest mass in a GRB jet fluid element is $\Gamma(1 + w)$, where $\Gamma \gg 1$ is the bulk Lorentz factor. The energy associated with radiation is $\Gamma w$, and the fraction of the total energy carried by radiation (i.e. the radiative efficiency) is

\be
\frac{L_\gamma}{L} = \frac{w}{1+w}.
\ee

\noindent In the downstream, we have $w_d = (e_d+p_d)/\rho_d = (4/3) \bar{\epsilon}_d (m_e n_\gamma / m_p n_p)$, or (using Equation~(\ref{eq:epsilon_d}) and assuming the upstream to be cold, $w_u \ll \gamma-1$)

\be
w_d = \frac{4}{3}(\gamma-1),
\ee

\begin{figure}
\includegraphics[width=\linewidth]{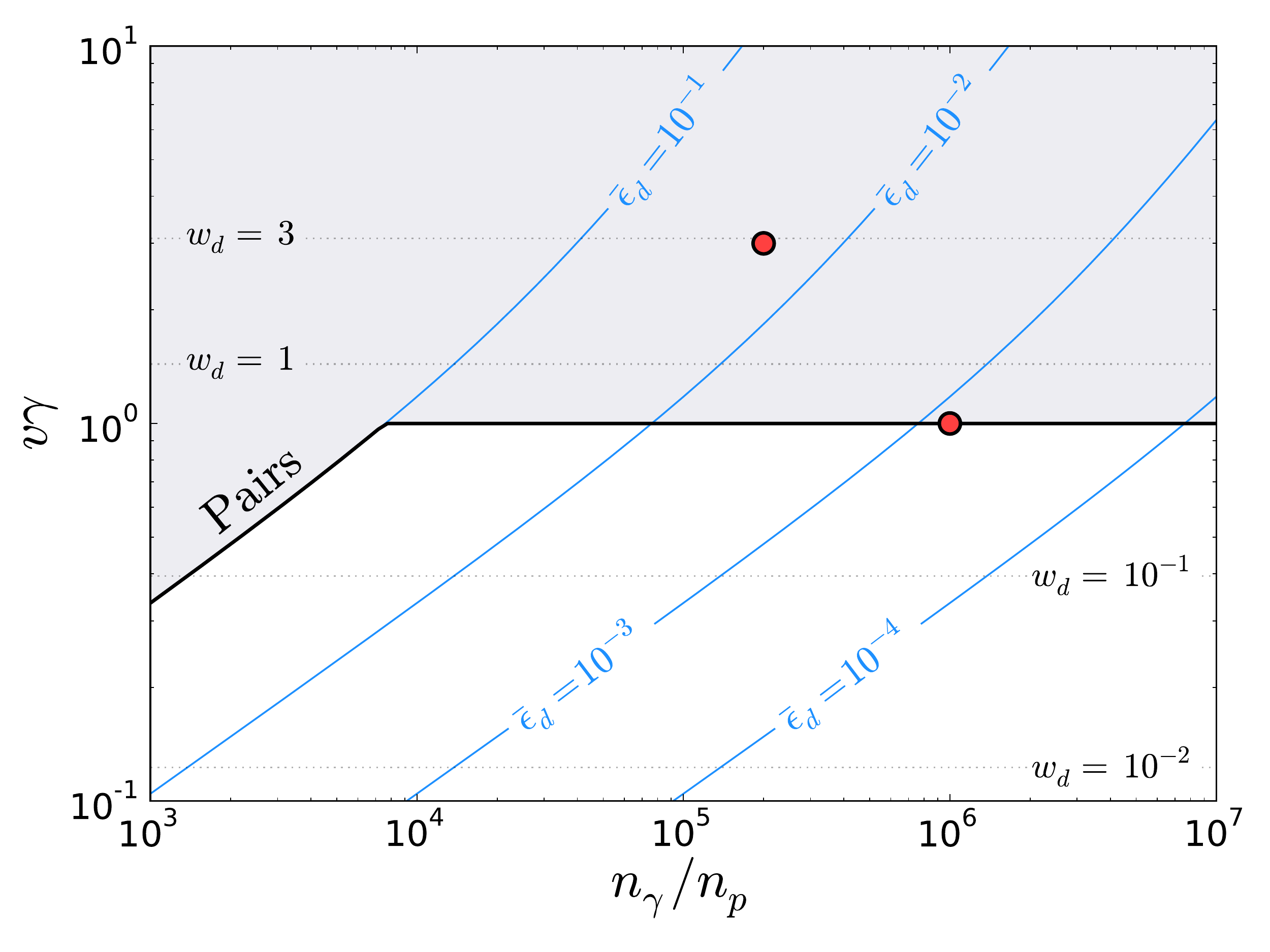}
\caption{Parameter space for photon-rich RMSs. The upstream flow is here assumed to be cold (i.e. $w_u \ll w_d$). The gray region indicates where pair production is expected to occur. The red dots show the parameters of the simulations presented in Section~\ref{sec:numerical}. (Note that there is a lower bound on $v\gamma$ for which shocks can occur for a given value of $w_u$; the sound speed is $c_s^2 = (1/3) w_u / (1 + w_u)$, and $v > c_s$ is required for a shock solution to exist.)}
\label{fig:parameter_space}
\end{figure}

\noindent Only sufficiently relativistic shocks, $v\gamma \simgt 0.4$, are capable of generating significant $L_\gamma/L \simgt 0.1$. We can then conclude that pair production is expected to occur in the RMSs which produce the most efficient GRB emission. The parameter space relevant to internal GRB shocks is shown in Figure~\ref{fig:parameter_space}. Shocks with $v\gamma \gtrsim 1$ (or very large average photon energies) are expected to produce pairs. Shocks capable of producing the observed GRB emission are expected to populate the approximate region of $1/3 \simlt v\gamma \simlt 3$ and $10^4 \simlt n_\gamma/n_p \simlt 10^6$, which would result in reasonable observed average photon energies, $\bar{E} \sim \Gamma\bar{\epsilon}_d m_e c^2 / (1+z)$, where $z$ is the GRB redshift.

\section{Numerical simulations: mildly relativistic shocks into photon-rich plasma}
\label{sec:numerical}

In this section we present results from four simulations. We consider two faster shocks with $v\gamma = 3$ and $n_\gamma/n_p = 2 \times 10^5$, and two slower shocks with $v\gamma = 1$ and $n_\gamma/n_p = 10^6$ (these sets of parameters are marked with two red dots in Figure~\ref{fig:parameter_space}). The faster shocks have $v\gamma \gtrsim 1$ and are therefore expected to produce large amounts of pairs, while the slower shocks should be close to the pair production boundary. The shocks run into an upstream which is either cold with $w_u = 3 \times 10^{-2}$, or warm with $w_u = 0.3$. Warmer upstreams ($w_u \lesssim 1$) are expected if the upstream material was recently heated, while fluid elements which were heated longer than a few expansion times ago will be colder ($w_u \ll 1$).

We consider homogeneous initial conditions across the whole grid for all runs. The simulation starts with constant values of the hydrodynamical parameters $v\gamma$, $\rho$ and $p$. Photons are injected across the whole grid with a (comoving) Wien spectrum, so that they are initially in local kinetic equilibrium with the electrons.

Below we present the steady-state shock structure (as seen in the downstream frame) and photon spectra for different initial conditions. The structure is plotted versus the ``original'' optical depth, defined as

\be
\tau_p(x) \equiv \int\limits_0^x \gamma n_p \sigma_\mathrm{T} \mathrm{d}x^\prime,
\ee

\noindent or the total optical depth which includes pairs,

\be
\tau_\pm(x) \equiv \int\limits_0^x Z_\pm \gamma n_p \sigma_\mathrm{T} \mathrm{d}x^\prime.
\ee

\noindent These definitions correspond to the original electron or pair columns of the fluid elements. The actual Thomson optical depth as seen by a photon also depends on the photon direction and the speed of the fluid elements.

\subsection {Initial shock evolution}

All runs follow qualitatively similar dynamical evolutions before settling into a steady-state. The initial conditions are set up so that the flow, which is initially moving to the left, immediately smashes into a lab frame wall (reflecting boundary) at the left end of the grid ($x = 0$). A hydrodynamical shock is formed at the left boundary, propagating in the rightward direction, while the downstream fluid becomes stationary in the lab frame ($v = 0$). The downstream region between the shock and the wall initially has a very small optical depth, so that photons are incapable of carrying the downstream pressure that is demanded by the shock jump conditions. The shock is therefore collisionless, and the downstream electron temperature is relativistic. Due to the large number of photons per electron, the electrons are quickly cooled in the downstream by a small fraction of the photons in the vicinity of the shock. The few photons which interact with the hot electrons quickly gain high energies. After a short time, the number of photons with energies $\epsilon \gtrsim 1$ is large enough so that their free paths to $\gamma\gamma$ collisions become smaller than the size of the downstream, triggering efficient pair production. The increase in the downstream optical depth causes more photons to scatter on the hot electrons, quickly cooling them and producing more pairs, until the downstream is optically thick, photons dominate the downstream energy density and electrons settle into kinetic equilibrium with radiation at the Compton temperature.

A fraction of the photons with $\epsilon \gtrsim 1$ leak ahead of the collisionless shock, sprinkling pairs into the upstream. As the upstream pair column becomes significant, photons can effectively ``grip'' the incoming upstream flow, and start gaining energy also by scattering back and forth across the collisionless shock. The rapid increase in photon pressure at the shock smears out the shock jump on a scale comparable to several photon mean free paths, smoothing out the collisionless shock and establishing proper radiation mediation. At this time, the shock has traversed a distance which corresponds to $\tau_p$ significantly less than unity. The shock settles into a steady-state after propagating for several upstream optical depths.

\subsection{Faster shock into cold upstream}

The fast shock simulations has parameters $v\gamma = 3$, $n_\gamma/n_p = 2 \times 10^5$ and a cold upstream with $w_u = 3 \times 10^{-2}$. These values correspond to an average upstream photon energy of $\bar{\epsilon}_u \sim 6.5 \times 10^{-4}$, as measured in the downstream frame. (The corresponding $\nu F_\nu$ peak of the Wien spectrum is $\epsilon_\mathrm{pk} \approx 4 \bar{\epsilon}_u$.) The speed difference between the upstream and downstream is large enough for pair production to become important, and the cold upstream ensures that essentially all downstream photon energy comes from the upstream proton kinetic energy.

\begin{figure}
\includegraphics[width=\linewidth]{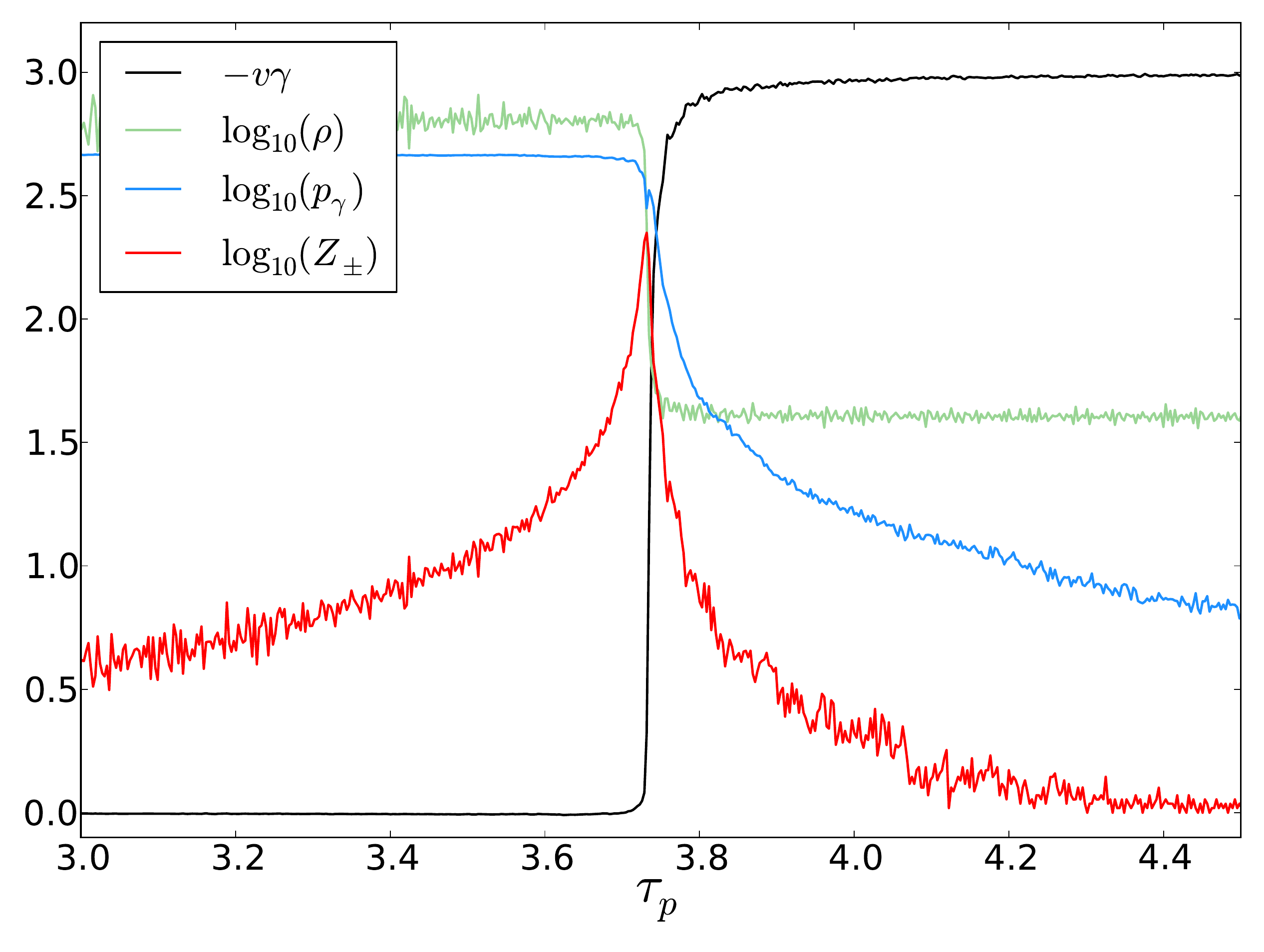}
\caption{Structure of the shock with parameters $v\gamma = 3$, $n_\gamma/n_p = 2 \times 10^5$ and $w_u = 3 \times 10^{-2}$. Profiles are shown of $v\gamma$ (downstream frame), $\rho$, $p_\gamma$ and $Z_\pm$ after steady-state is reached. The horizontal coordinate, $\tau_p$, is the optical depth associated to the ``original'' electrons. The upstream (right part of the figure) is flowing to the left. (Note that the absolute values of $p_\gamma$ and $\rho$ carry no significance for the planar shock problem, but the ratio $p_\gamma/\rho$ does; an upstream value of $\rho = 40$ g cm$^{-3}$ was used for this plot.)}
\label{fig:hydro_parameters_fc}
\end{figure}

The steady-state shock structure is shown in Figure~\ref{fig:hydro_parameters_fc} as a function of the original optical depth, $\tau_p$. A photon precursor is leaking into the upstream, pre-heating the electrons and sprinkling pairs ahead of the shock. The photon pressure gradient increases toward the shock, decelerating the incoming upstream flow. The pair multiplicity peaks immediately behind the shock due to pair production and annihilation balance, with $Z_\pm \approx 225$ as its largest value. The large value of $Z_\pm$ decreases the photon mean free path by about the same factor, causing the shock transition to occur on a very short spatial length scale.

\begin{figure}
\includegraphics[width=\linewidth]{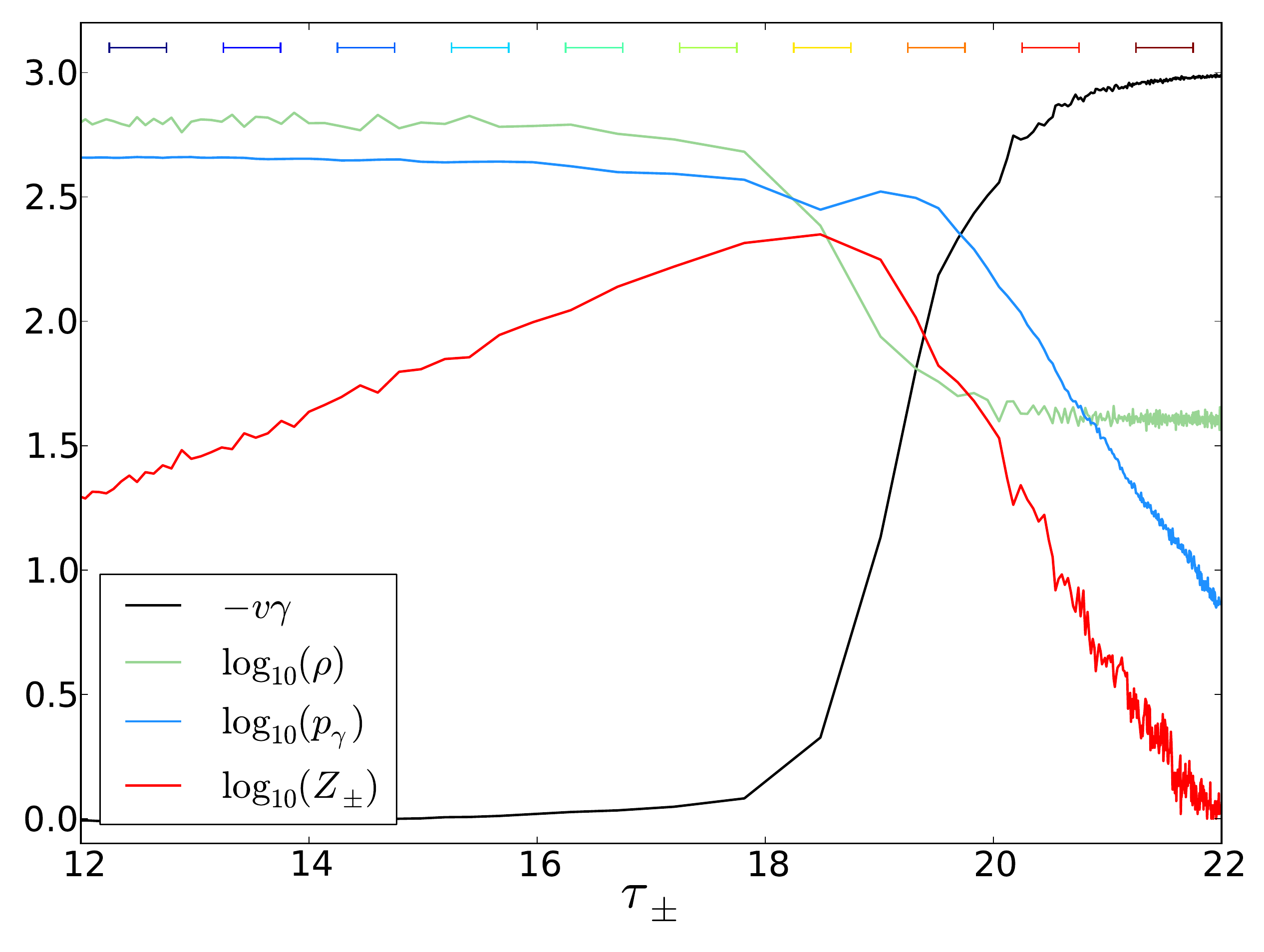}
\caption{The same snapshot of $v\gamma$, $\rho$, $p_\gamma$ and $Z_\pm$ as in Figure~\ref{fig:hydro_parameters_fc} ($v\gamma = 3$, $n_\gamma/n_p = 2 \times 10^5$, $w_u = 3 \times 10^{-2}$), but here shown as functions of the total optical depth $\tau_\pm$, which includes the (dominant) contribution from pairs. The colored bars at the top of the figure define regions within the flow. The photon spectra within these regions are plotted with the corresponding colors in Figures~\ref{fig:spectra_fc} and \ref{fig:spectra_fc_nuFnu}.}
\label{fig:hydro_parameters_fc_pm}
\end{figure}

The detailed shock structure is more clearly visible in Figure~\ref{fig:hydro_parameters_fc_pm}, which shows the same shock profile as a function of the total optical depth, $\tau_\pm$. As expected, the shock transition region is smeared out over a few optical depths. The photon precursor pressure and the pair multiplicity are decreasing roughly exponentially toward the upstream, ahead of the shock. The shock structure as shown in Figure~\ref{fig:hydro_parameters_fc_pm} is similar to the structure of shocks with $v\gamma \lesssim 1$, which do not produce pairs.

\begin{figure}
\includegraphics[width=\linewidth]{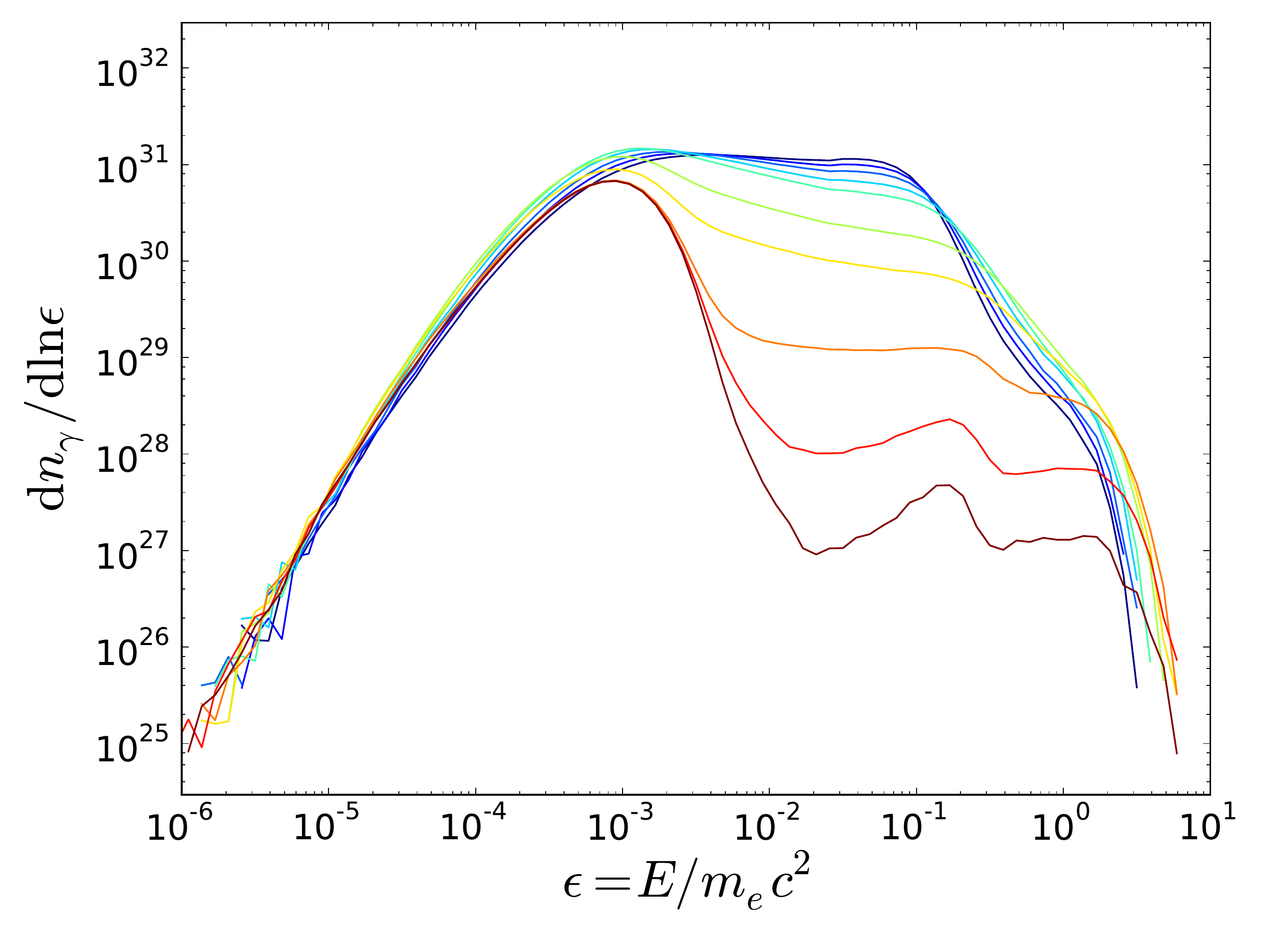}
\caption{Steady-state downstream frame photon number spectra at different locations within the shock with parameters $v\gamma = 3$, $n_\gamma/n_p = 2 \times 10^5$ and $w_u = 3 \times 10^{-2}$. The line colors correspond to the locations indicated by the colored bars at the top of Figure~\ref{fig:hydro_parameters_fc_pm}.}
\label{fig:spectra_fc}
\end{figure}

The lab frame (i.e. downstream frame) photon number spectra at different locations within the shock structure are shown in Figure~\ref{fig:spectra_fc}. The spectra are collected at locations which are separated by an optical depth of unity, as indicated by the colored bars in Figure~\ref{fig:hydro_parameters_fc_pm}. The upstream spectra (dark red and red) show the Wien spectrum shape at energies of $\epsilon \sim 2 \times 10^{-4}$, and a precursor of high-energy photons. The photon number spectrum per logarithmic interval in energy is roughly flat at high energies, and the photon spectrum above $1/\epsilon_{max}$ is affected by $\gamma\gamma$-absorption. The spectrum at the base of the shock transition (yellow) is essentially a power law extending from $\epsilon_u$, which starts softening around $\epsilon \gtrsim 10^{-1}$. A significant fraction ($\sim 10^{-2}$) of the photons inside the shock have energies above $\epsilon = 1$, giving rise to strong pair production inside the shock. The downstream spectra (green to blue) show the gradual process of ``thermalization'' toward a Wien spectrum. The spectrum evolves more quickly at high energies, because of the energy transfer to electrons through recoil in scattering.

\begin{figure}
\includegraphics[width=\linewidth]{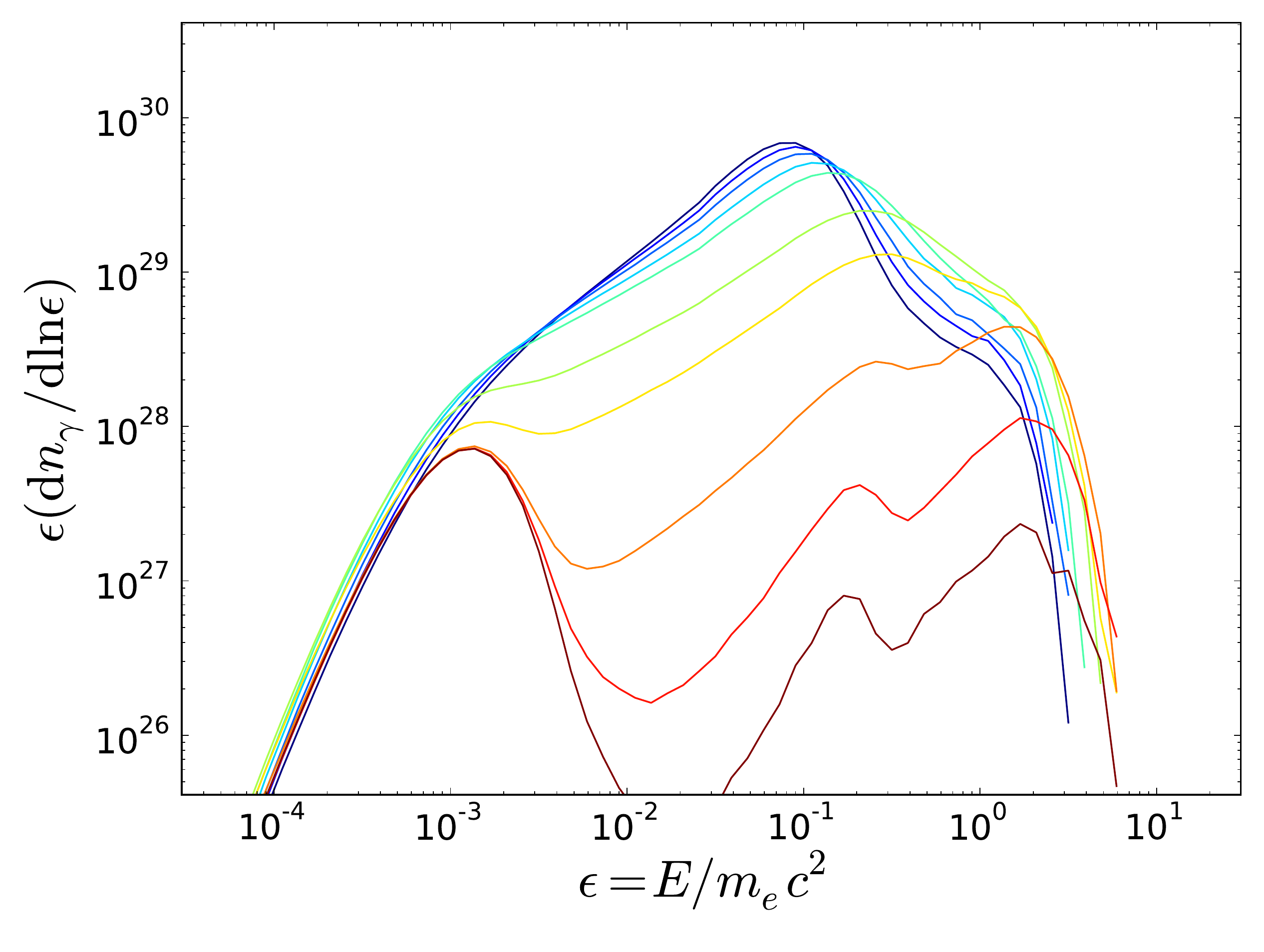}
\caption{Same as Figure~\ref{fig:spectra_fc}, but now showing total energy density per logarithmic interval in energy (i.e. $\nu F_\nu$) and zoomed in around the spectral peak.}
\label{fig:spectra_fc_nuFnu}
\end{figure}

Figure~\ref{fig:spectra_fc_nuFnu} shows the same spectra as Figure~\ref{fig:spectra_fc}, but zoomed in around the spectral peak and shown in the $\nu F_\nu$ representation. The peak is shifting from larger to smaller energies because of efficient recoil losses, transferring the energy to low energy photons.

\subsection{Faster shock into warm upstream}

This simulation has the same parameters as the previous simulation ($v\gamma = 3$, $n_\gamma/n_p = 2 \times 10^5$), but the upstream is warmer with $w_u = 0.3$, corresponding to $\bar{\epsilon}_u \approx 6.5 \times 10^{-3}$ as measured in the downstream frame. This implies a smaller energy amplification factor of photons crossing the shock. The shock structure as a function of $\tau_\pm$ is very similar to the previous simulation. However, the higher $\bar{\epsilon}_u$ leads to a somewhat different radiation spectrum.

\begin{figure}
\includegraphics[width=\linewidth]{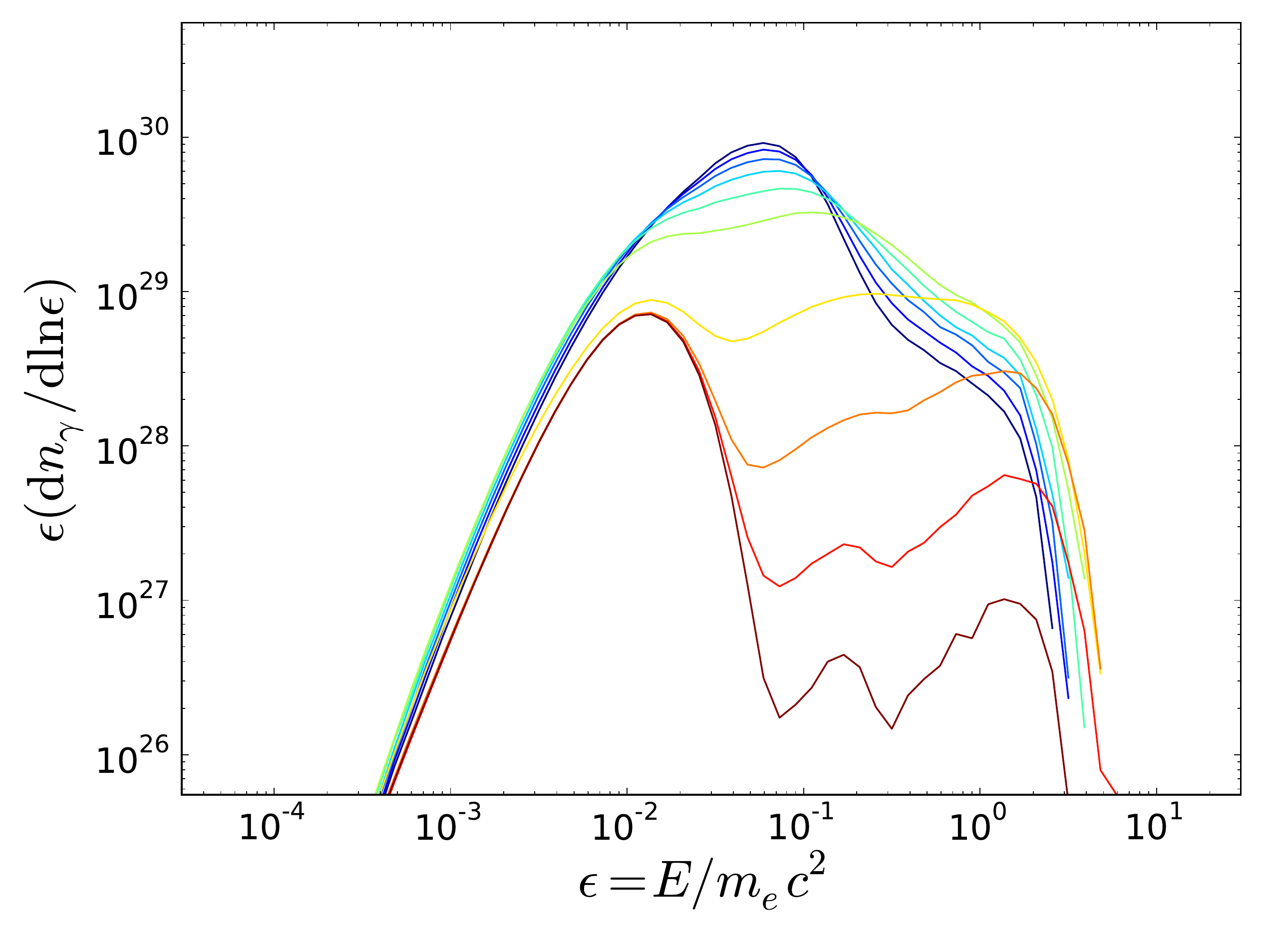}
\caption{Steady-state downstream frame spectra at different locations within the shock with parameters $v\gamma = 3$, $n_\gamma/n_p = 2 \times 10^5$ and $w_u = 3 \times 10^{-1}$. The spectral are taken at the same relative locations within the shock as in Figure~\ref{fig:spectra_fc_nuFnu}.}
\label{fig:spectra_fw_nuFnu}
\end{figure}

The $\nu F_\nu$ shock spectra are shown in Figure~\ref{fig:spectra_fw_nuFnu}, as a function of location within the shock. The colors correspond to the same locations as for the fast shock into the cold upstream. The smaller amplification factor of photon energies leads to the softer spectrum inside the shock (yellow curves). This reduces the number of photons with $\epsilon>1$ and thus decreases pair loading to $Z_\pm\approx 100$.


\subsection{Slower shock into cold upstream}

\begin{figure}
\includegraphics[width=\linewidth]{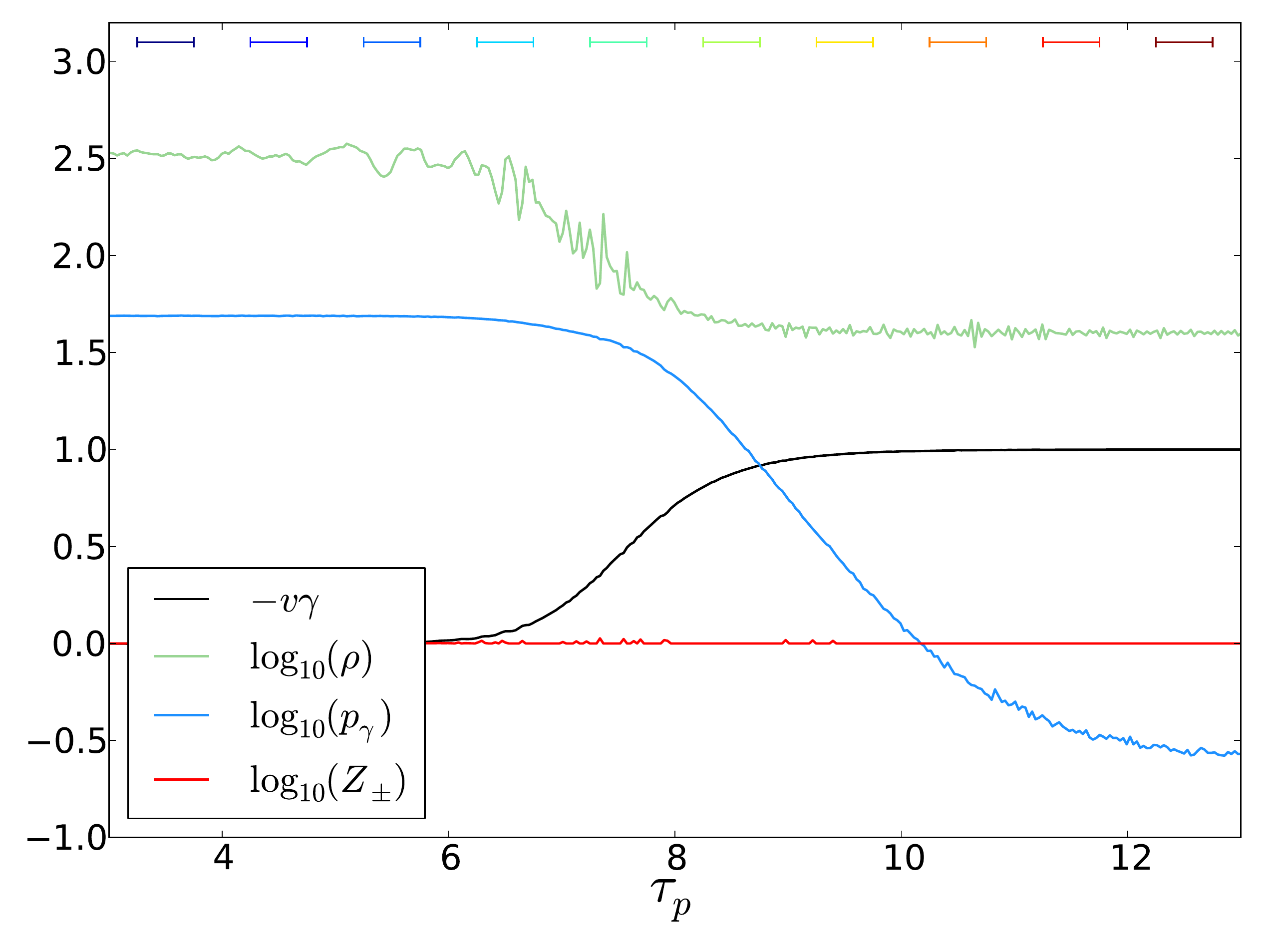}
\caption{Structure of the shock with parameters $v\gamma = 1$, $n_\gamma/n_p = 10^6$ and $w_u = 3 \times 10^{-2}$. Profiles are shown of $v\gamma$ (downstream frame), $\rho$, $p_\gamma$ and $Z_\pm$ after steady-state is reached. The colored bars at the top of the figure define regions within the flow. The photon spectra within these regions are plotted with the corresponding colors in Figure~\ref{fig:spectra_sc_nuFnu}.}
\label{fig:hydro_parameters_sc}
\end{figure}

The slower shock has an upstream speed corresponding to $v\gamma = 1$, which is right on the expected boundary for pair production. We consider $n_\gamma/n_p = 10^6$ and the cold upstream has $w_u = 3 \times 10^{-2}$, corresponding to $\bar{\epsilon}_u \approx 3.0 \times 10^{-5}$. The shock structure is shown in Figure~\ref{fig:hydro_parameters_sc}. The shock transition occurs over a few optical depths. The pair multiplicity is equal to unity, although tiny ``bumps'' can be seen in the red $Z_\pm$ line, indicating that these shock parameters are \textit{just} below the threshold for increasing the pair multiplicity.

\begin{figure}
\includegraphics[width=\linewidth]{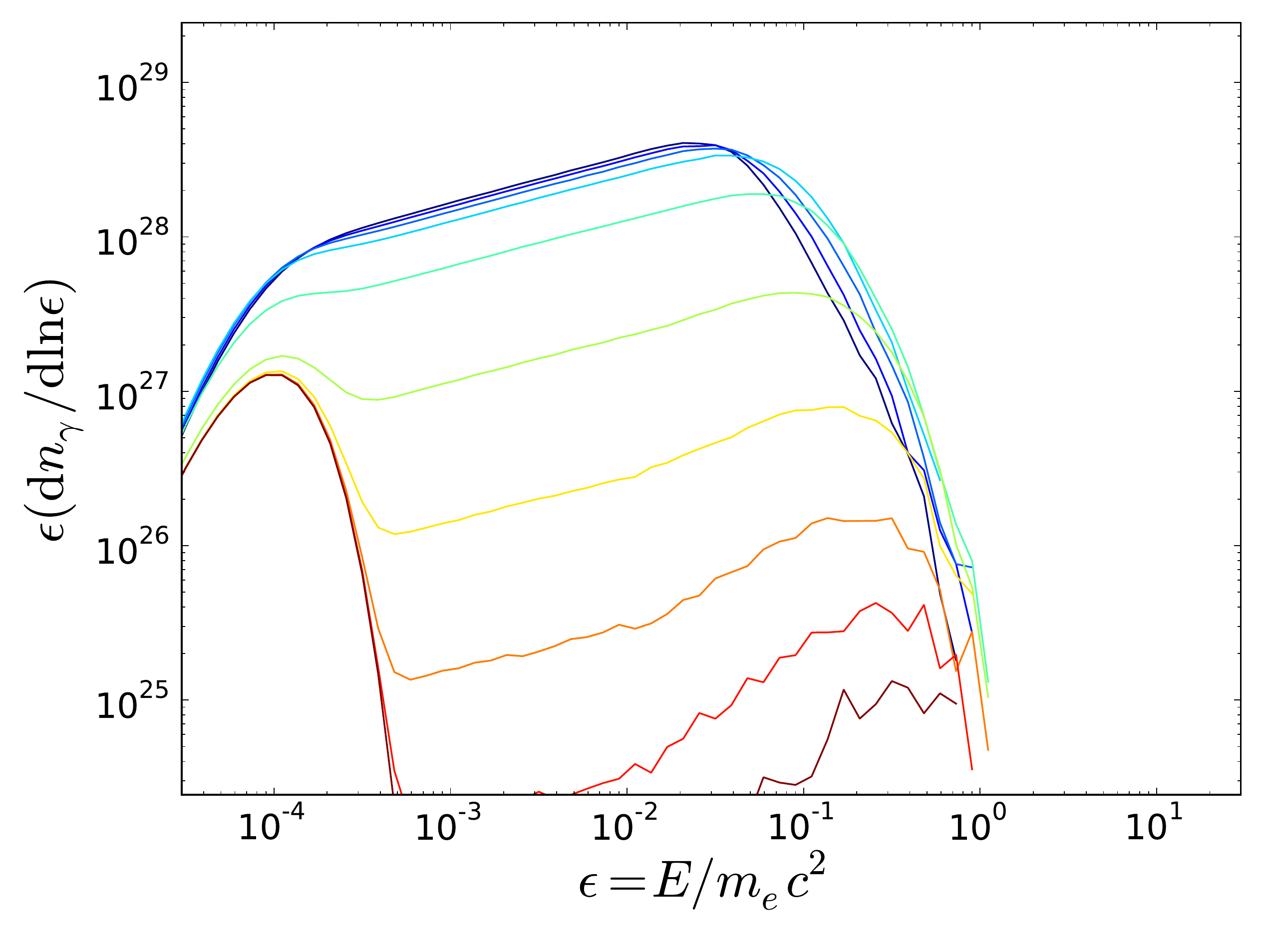}
\caption{Steady-state downstream frame spectra at different locations within the shock with parameters $v\gamma = 1$, $n_\gamma/n_p = 10^6$ and $w_u = 3 \times 10^{-2}$. The line colors correspond to the locations indicated by the colored bars at the top of Figure~\ref{fig:hydro_parameters_sc}.}
\label{fig:spectra_sc_nuFnu}
\end{figure}

Figure~\ref{fig:spectra_sc_nuFnu} shows the $\nu F_\nu$ spectrum at different locations within the shock. As before, the locations are indicated in Figure~\ref{fig:hydro_parameters_sc}. The upstream photon energy is very small, and the spectrum at the shock base (light blue) is a perfect power law for several decades in energy, extending up to $\epsilon \sim 10^{-1}$. The precursor hardens slightly toward the upstream due to the increased mean free path for higher energy photons. The fraction of photons inside the shock with energy $\epsilon \gtrsim 1$ is less than $10^{-6}$, which is marginal for not increasing the pair multiplicity above $Z_\pm = 1$.

\subsection{Slower shock into warm upstream}

\begin{figure}
\includegraphics[width=\linewidth]{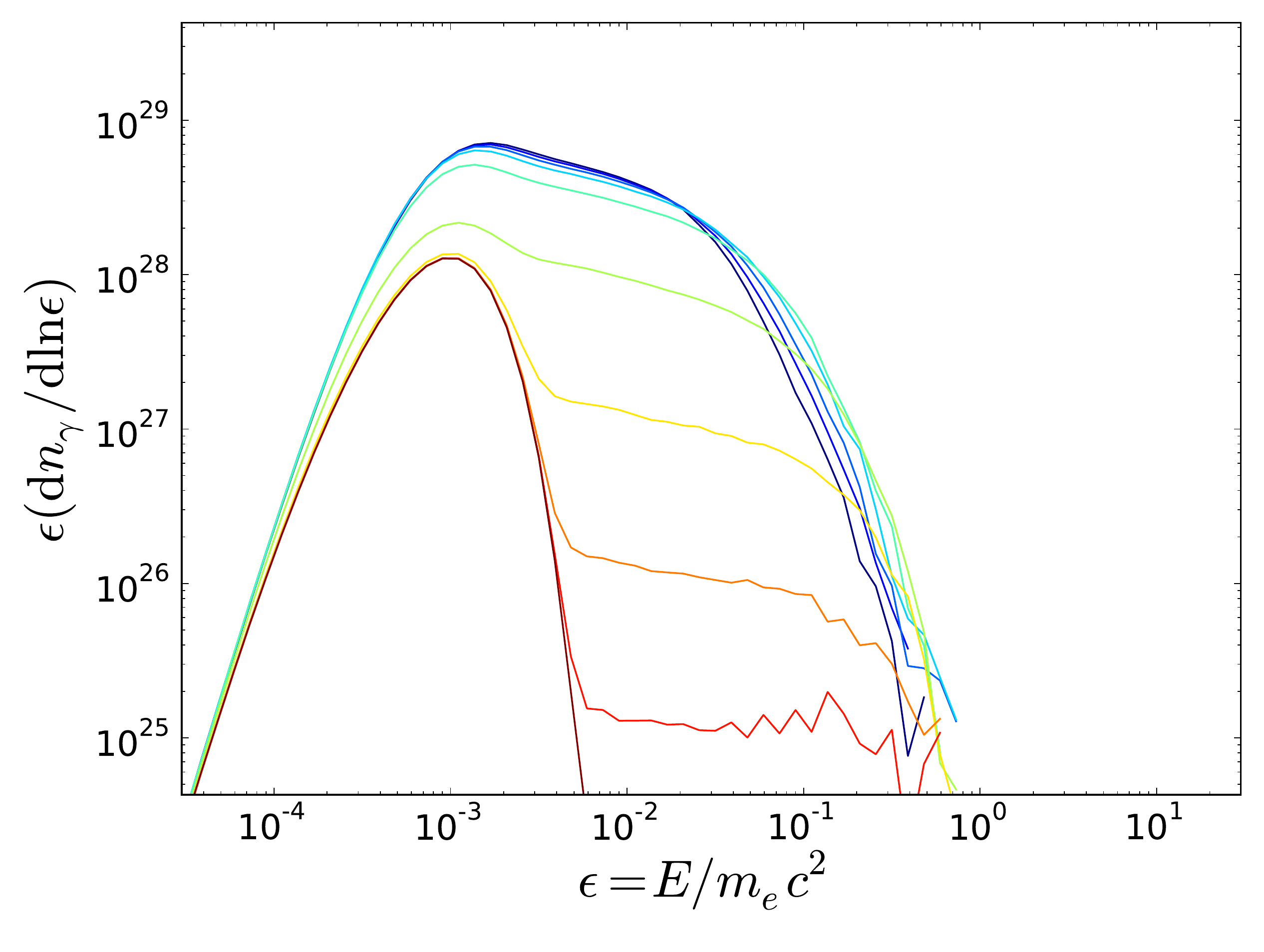}
\caption{Steady-state downstream frame spectra at different locations within the shock with parameters $v\gamma = 1$, $n_\gamma/n_p = 10^6$ and $w_u = 3 \times 10^{-1}$, taken at the same relative locations within the shock as for Figure~\ref{fig:spectra_sc_nuFnu}.}
\label{fig:spectra_sw_nuFnu}
\end{figure}

Here we used the same parameters for the upstream speed and photon number as for the previous simulation ($v\gamma = 1$ and $n_\gamma/n_p = 10^6$), but the upstream is warmer with $w_u = 0.3$, corresponding to $\bar{\epsilon}_u \approx 3.0 \times 10^{-4}$. The hydrodynamic shock structure is the same as in the cold simulation. The $\nu F_\nu$ spectrum is shown in Figure~\ref{fig:spectra_sw_nuFnu}. Just as for the faster shocks, a warmer upstream leads to a softer power law spectrum, since the shock must arrange itself to give the photons a smaller energy amplification factor. The fraction of photons with energy $\epsilon \gtrsim 1$ in the shock is well below $10^{-6}$, so that $Z_\pm = 1$ throughout the shock.

\section{Discussion}
\label{sec:discussion}

\subsection{Summary of the main results}
\label{ssec:summary}

In this paper we have presented a time dependent, special relativistic radiation hydrodynamics code. The code is designed specifically for simulating radiation mediated shocks (RMSs), and incorporates full Klein-Nishina scattering and $\gamma\gamma$-pair production. We have used our code to calculate the fully self-consistent RMS structure in media where upstream photon advection is the main photon source; this is the case for RMSs inside (unmagnetized) GRB jets.

We have simulated RMSs of various speeds and upstream conditions. The shocks were allowed to propagate until they settled into steady-states, after which the shock structure has been examined. In particular, the photon spectra and the pair-to-proton ratio $Z_\pm$ have been analyzed as a function of location within the shock transition.

RMSs heat photons via the first order Fermi process, producing power law photon spectra within the shock transition region. The largest photon energy inside a non-relativistic shock is $\epsilon_{max} = E_{max}/m_e c^2 \sim v^2$, where $v$ is the upstream speed as measured in the downstream (in units of the speed of light), and the power law index depends on the shock speed, the upstream photon-to-proton ratio and the average upstream photon energy.

The photon spectra inside non-relativistic RMSs are pure power laws, extending from the typical upstream photon energy up to $\epsilon_{max}$. Shocks with $v\gamma \lesssim 1$ do not produce pairs, as they do not heat photons up to the electron rest mass. On the other hand, shocks with $v\gamma \gtrsim 1$ heat photons to $\epsilon \gtrsim 1$ and are therefore strong producers of pairs, with typical values of $Z_\pm \gtrsim 10^2$ inside the RMS transition. The optical depth of the plasma is increased by a factor $\sim Z_\pm$, and the spatial width of the RMS is correspondingly decreased by the same factor. The power law photon spectra inside such RMSs curve downward at $\epsilon \sim 10^{-1}$ due to Klein-Nishina effects, and are affected by $\gamma\gamma$-annihilation at $\epsilon \sim 1$. The pairs annihilate behind the RMS, and the photon spectra gradually thermalize toward the downstream. In the absence of photon production processes, the spectra approach the Wien spectrum at an optical depth $\tau_{th} \sim 1/3 \bar{\epsilon}_d$ behind the shock, where $\bar{\epsilon}_d$ is the average downstream photon energy.

\subsection{The ``single plasma'' assumption}
\label{ssec:single_plasma}

Our implementation of the hydrodynamics assumes that the plasma behaves as a single fluid, so that a single speed and temperature can be defined for each fluid element. In a RMS, the photons interact with electrons (or positrons), and the electrons are subsequently coupled to the protons. The coupling (i.e. isotropization of the electrons) is maintained on length scales of order the plasma skin depth, which is always much shorter than the photon mean free path. 


The time for charged particles to relax to a Maxwellian distribution is set by Coulomb collisions. In the limit $\theta \ll 1$, which is valid for the shocks considered here, the pair relaxation time $t_\pm$ is \citep{Stepney:1983}

\be
\frac{t_\pm}{t_{sc}} \approx \frac{2\pi^{1/2}}{\ln\Lambda} \theta^{3/2} \approx 0.17 \theta^{3/2} \ll 1,
\label{eq:t_pm}
\ee

\noindent where $t_{sc} = (Z_\pm n_p \sigma_\mathrm{T} c)^{-1}$ is the local photon-electron scattering time and $\ln\Lambda \approx 20$ is the Coulomb logarithm. The time for electron-electron relaxation is twice that for electron-positron relaxation. The times and length scales in a RMS (in the absence of a subshock) are set by $t_{sc}$, and Equation~(\ref{eq:t_pm}) implies that electrons (and pairs) maintain a local Maxwellian distribution.


The timescale for electron-proton relaxation is longer,

\be
\frac{t_{ep}}{t_{sc}} \approx \sqrt{\frac{\pi}{2}} \frac{Z_\pm m_p}{m_e \ln\Lambda} \left(\theta_e + \frac{m_e}{m_p}\theta_p\right)^{3/2} \approx 120 Z_\pm \theta_e^{3/2},
\ee

\noindent so that electrons may not have time to exchange energy with the protons throughout the shock, depending on the shock parameters. On the other hand, the heat capacity of the protons is extremely small compared with that of radiation (due to the huge number of photons per proton), and the exact details of their internal energy is unimportant for the shock problem. We therefore conclude that the ``single plasma'' assumption is valid for RMSs which propagate into unmagnetized, photon-rich upstreams.

\subsection{Neutrons}
\label{ssec:neutrons}

GRB jets can have a significant neutron component \citep{DerEtAl:1999, Bel:2003}. Neutron mediated shock waves were discussed by B17. The cross-section for nuclear collisions is smaller than the Thomson cross-section, $\sigma_n/\sigma_\mathrm{T} \sim 1/20$, and the neutron mean free path is therefore

\be
\lambda_n/\lambda \sim 20 Z_\pm/(1+Z_n),
\ee

\noindent where $Z_n \equiv n_n/n_p$ is the ratio of neutrons to protons in the flow. The neutron mean free path is larger than the photon mean free path (unless the flow is very neutron rich with $Z_n > 20 Z_\pm$), and the RMS can therefore exist as a subshock inside a broader neutron mediated shock. If the neutron component is small, $Z_n \ll 1$, then the neutron shock acts as a weak precursor to the RMS, and the RMS dissipates most of the energy.

In this work we considered a neutron-poor plasma ($Z_n \ll 1$). In principle, neutrons could be simulated as Monte Carlo particles along with the photons, although additional numerical challenges are introduced. Mildly relativistic neutron-proton collisions generate pions, which quickly decay into relativistic ($\gamma_e \approx m_\pi/m_e \sim 300$) electron-positron pairs \citep{DerEtAl:1999}. The relativistic pairs subsequently launch a pair cascade \citep{Bel:2010, VurEtAl:2011}. The assumption of thermal electrons is not valid in this case.

\subsection{Observations of RMS spectra}

The spectra presented in this work are the steady-state shock spectra (as viewed from the downstream). RMSs can only attain steady-state as long as the local optical depth is large (or, equivalently, the scattering time is smaller than the jet expansion time). There is then a qualitative difference between ``deep'' and ``shallow'' shocks. Deep shocks dissipate most of their energy at $\tau \gg 1$, while shallow shocks dissipate most of their energy at about $\tau \lesssim 10$. Deep shocks are effectively planar. Shocked fluid elements continue to expand (and perhaps will be shocked again) as they approach the photosphere. The shock-amplified photons continue to scatter until they reach the photosphere and start streaming freely. Scattering tends to ``thermalize'' the photon spectrum, and the combination of scattering and expansion leads to adiabatic energy losses. All pairs have time to annihilate for shocks occuring at $\tau \gg 1$. Thus the escaping spectrum from a deep shock is expected to appear like a partially thermalized RMS spectrum which has suffered adiabatic energy losses. A Wien spectrum will be formed if the shock occured well inside the Wien zone \citep{Bel:2013}, where the thermal Compton $y$-parameter is large, $y \sim \tau \bar{\epsilon}_d \gg 1$. Furthermore, the observed spectrum is necessarily integrated over the shock downstream due to the short time variability of the flow \citep{Lev:2012}, and also likely composed of time integration over several shock episodes \citep{KerLev:2014}.

Shallow shocks can be significantly different and will be studied in a separate paper (C. Lundman and A. M. Beloborodov, in preparation). The planar approximation is expected to break down when the local scattering time becomes comparable to the expansion time (roughly at $\tau \lesssim 10$). A non-planar geometry causes the local comoving radiation intensity to become beamed along the local flow direction \citep{Bel:2011}. The long scattering time makes photons less efficient in mediating the shock, and the flow is expected to try to develop a collisionless subshock as the shock ``breaks out'' of the photosphere. B17 pointed out that the shock will ``dress'' itself in pairs, maintaining a significant optical depth even far outside the nominal photosphere of the GRB jet. Non-planar, time dependent numerical simulations are needed to fully assess the details of GRB shock breakouts.

\acknowledgments
The authors would like to thank Hirotaka Ito for useful discussions. CL acknowledges the Swedish Research Council for financial support. AMB is supported by NSF grant AST-1412485, NASA grant NNX15AE26G, and a grant from the Simons Foundation (\#446228, Andrei Beloborodov). IV acknowledges support from the Estonian Research Council grant PUT1112.

\bibliographystyle{apj}
\bibliography{refgrb}


\appendix

\section{Lagrangian hydrodynamic equations}
\label{app:derivation}

We here specialize Equations~(\ref{eq:energy_momentum_cons}), (\ref{eq:mass_cons}) and (\ref{eq:pair_equation}) to planar, one-dimensional flows. We denote the spatial coordinate as $x$. The equations for the conservation of proton number (Equation~(\ref{eq:mass_cons})), energy and momentum (Equation~(\ref{eq:energy_momentum_cons})) then become

\be
(\Gamma\rhop)_{,t} + (\Gamma \beta \rhop)_{,x} = 0,
\label{eq:mass}
\ee

\be
(\Gamma^2 h - p)_{,t} + (\Gamma^2 h \beta)_{,x} = G^0,
\label{eq:energy}
\ee

\noindent and

\be
(\Gamma^2 h \beta)_{,t} + (\Gamma^2 h \beta^2 + p)_{,x} = G^1,
\label{eq:momentum}
\ee

\noindent respectively, where $h \equiv \rho + e + p$.

We now define Lagrangian coordinates, for which the partial time derivative is taken for a given fluid element as opposed to at a fixed spatial coordinate; $\p/\p t \rightarrow \p/\p t - \beta\p/\p x$ and $\p/\p x \rightarrow \p/\p x$. The spatial coordinate is then replaced by the proton mass coordinate $m$, defined as

\be
m \equiv \int\limits_{x_{min}}^x \Gamma\rhop \, \D x^\prime,
\ee

\noindent so that $\p/\p x \rightarrow \Gamma\rhop \, \p/\p m$. Re-writing Equations~(\ref{eq:mass}), (\ref{eq:energy}) and (\ref{eq:momentum}) in terms of the new coordinates gives

\be
(\Gamma\rhop)_{,t} + (\Gamma\rhop)^2 \beta_{,m} = 0,
\label{eq:mass_2}
\ee

\be
(\Gamma^2 h - p)_{,t} + \Gamma\rhop \left[\beta p_{,m} + \Gamma^2 h \beta_{,m}\right] = G^0
\label{eq:energy_2}
\ee

\noindent and

\be
(\Gamma^2 \beta h)_{,t} + \Gamma\rhop \left[p_{,m} + \Gamma^2 \beta h \beta_{,m}\right] = G^1.
\label{eq:momentum_2}
\ee

\noindent Finally, we introduce the lab frame volume, energy, and momentum per proton rest mass $V_p$, $E_p$, and $S_p$ as new variables,

\be
V_p \equiv \frac{1}{\Gamma \rhop},
\ee

\be
E_p \equiv \frac{\Gamma^2 h - p}{\Gamma \rhop},
\ee

\be
S_p \equiv \frac{\Gamma^2 \beta h}{\Gamma \rhop}.
\ee

\noindent Re-writing Equations~(\ref{eq:mass_2}), (\ref{eq:energy_2}) and (\ref{eq:momentum_2}), we obtain the one-dimensional, planar equations of special relativistic Lagrangian hydrodynamics with energy and momentum source terms,

\be
(V_p)_{,t} - \beta_{,m} = 0,
\ee

\be
(E_p)_{,t} + (p \beta)_{,m} = V_p G^0
\ee

\noindent and

\be
(S_p)_{,t} + p_{,m} = V_p G^1.
\ee

The Lagrangian equation for the pair loading is found by noting that $Z_\pm \equiv n_\pm/n_p$ and $(n_p u^\alpha)_{;\alpha} = 0$, so that $(n_\pm u^\alpha)_{;\alpha} = n_p u^\alpha (Z_\pm)_{;\alpha} = \dot{n}_\pm$. Changing to Lagrangian coordinates, $\partial_t \rightarrow \partial_t - \beta\partial_x$, we find the pair loading equation,

\be
(Z_\pm)_{,t} = m_p V_p \dot{n}_\pm.
\ee

\end{document}